\pdfoutput=1
\documentclass[runningheads]{llncs}
\usepackage[T1]{fontenc}
\usepackage[utf8]{inputenc}
\usepackage{graphicx}
\usepackage{amsmath,amssymb}
\usepackage{xurl}
\usepackage[hidelinks]{hyperref}
\usepackage{xcolor}
\usepackage{tikz}
\usepackage{array}
\newcolumntype{L}[1]{>{\raggedright\arraybackslash}p{#1}}
\usetikzlibrary{arrows.meta,positioning,fit}
\newcommand{\runin}[1]{\par\smallskip\noindent\textbf{\textit{#1}}}
\spnewtheorem{scenario}{Scenario}{\itshape}{\rmfamily}
\begin{document}

\title{Model Weights, Airlocks and Sandboxes\\
\large  A Call for Technical Validation of Re-identification Tests under the EDPB Anonymisation Guidelines}
\titlerunning{Model Weights, Airlocks and Sandboxes}

\author{Rickard Br\"annvall
\and Henrik Forsgren \and
Erik Ylip\"a\"a \and \\
Aslak Felin \and
Michael Popoff \and
Susanne Stenberg}
\authorrunning{R. Br\"annvall et al.}
\institute{RISE Research Institutes of Sweden\\
\email{rickard.brannvall@ri.se}}

\maketitle

\begin{abstract}
The Guidelines 02/2026 on anonymisation of the European Data Protection Board (EDPB), published for consultation in July 2026, assess anonymity relative to the entities that may access data.
They bring AI models into scope and call for testing, but do not say how a re-identification test should be designed or validated, or how far its result counts as evidence.
Building on our reply to the consultation, we describe a framework based on a secure processing environment whose outputs are reviewed before release (an airlock), and distinguish forms of release and three types of evidence: formal, empirical and indirect.
We raise seven open questions, each with a scenario, and examine their technical side in three settings: the release of model weights (the trained parameters of a machine learning model), review at an airlock, and regulatory sandboxes, in which an authority supervises development and testing.
In each setting the assessment depends on how far a test result can be relied on, since a failed re-identification attempt may mean only that the test was too weak.
Drawing on measurement practice in other regulated fields, we call for re-identification tests to be validated as measurement methods, with stated detection limit, bias and uncertainty. 
\end{abstract}

\section{Introduction}\label{sec:n-intro}
A decision to release data or a trained model as anonymous rests on evidence that individuals cannot be re-identified.
The European Data Protection Board (EDPB) adopted Guidelines 02/2026 on anonymisation for public consultation on 7 July 2026~\cite{edpb2026anon}; unless stated otherwise, paragraph numbers in this paper refer to them.
The guidelines assess anonymity relative to the entities that may access the data and the means they are reasonably likely to use.
AI models fall within the same framework.
The assessment arises when an organisation considers sharing the weights of a model trained on personal data, when the operator of a secure processing environment reviews which outputs may leave it, and in those regulatory sandboxes where personal data is processed under the supervision of an authority.

Testing is central to such assessments.
The guidelines ask for documentation of the anonymisation process, including the testing (\S~41), and describe a test for whether an inference relies on an individual's data (\S~71b).
They do not say how a test should be designed, what its result shows, how its uncertainty should be reported or how the test method should be validated, and they refer to none of the existing standards on privacy risk and de-identification.
The need for such guidance has been noted elsewhere.
In the 2025 consultation on the EDPB's pseudonymisation guidelines~\cite{edpb2025pseudo}, by our count at least 13 of the 62 published replies asked for a risk assessment methodology~\cite{edpb2025consultation}.
A report for the EDPB on privacy risks in large language models calls for robust testing and validation processes~\cite{barbera2025llm}, a Swedish government inquiry on privacy-enhancing technologies in public administration finds that the lack of general guidelines for testing resistance to re-identification creates uncertainty~\cite[p.~70]{sou2026_44}, and practitioners report a lack of coherent guidance on when anonymity claims hold~\cite{lestyan2025washing}.

A failed re-identification attempt may mean that the data is well protected, or that the test was too weak to detect what it was looking for.
Research on empirical privacy evaluation finds that tests which miss the most vulnerable records or rely on biased evaluation data can understate or overstate the risk~\cite{aerni2024misleading,das2025blind}.
Other fields that base decisions on measurements validate their test methods on cases where the outcome is known, and their standards offer concepts for stating how far a result can be relied on.
We argue that re-identification tests should be treated in the same way.

This paper develops our reply to the consultation~\cite{rise2026reply} and makes four contributions:
\begin{enumerate}
\item A framework that places the two approaches of the guidelines in a setting used in practice, a secure processing environment with a release review (an airlock), and distinguishes the forms a release can take and three types of evidence. It is intended to give the legal and the technical side common terms for what is released, to whom, and on what evidence (Sect.~\ref{sec:n-framework}).
\item Seven open questions on how the guidelines apply to re-identification tests, AI models and secure environments, each with a scenario. They concern points where guidance or further research may be needed (Sect.~\ref{sec:n-points}).
\item A discussion of the technical side of these questions for model weights, airlocks and regulatory sandboxes, which examines what an answer would depend on in each setting (Sect.~\ref{sec:n-discussion}).
\item A call for re-identification tests to be validated as measurement methods, with an outline of what validation involves and what a test report could contain (Sect.~\ref{sec:n-discussion}, Method validation).
\end{enumerate}

We write from a technical perspective.
The questions we raise bear on the law, but we leave their assessment to others and aim to contribute technical knowledge to the legal discussion.
Our focus is on what can be inferred about individuals from aggregate outputs and trained models, which the guidelines treat under their No Inference criterion.
We refer to the guidelines in the version published for consultation, which may change.

\section{Background}\label{sec:n-background}
Under the GDPR, data is anonymous if no individual is identifiable by means reasonably likely to be used (Recital~26).
The EDPB Guidelines 02/2026 set out a framework for this assessment~\cite{edpb2026anon}.
The guidelines use \emph{given data} for the data being assessed for anonymity and \emph{entity} for a party in whatever role, such as a potential controller, processor or third party (\S~5).
An entity's \emph{means} are understood broadly, from reading a document to running complex analyses (\S~28).
Whether they are reasonably likely to be used depends on objective factors, among them the available technology and the \emph{additional information} the entity can obtain, that is, information beyond the given data that would allow individuals to be identified (\S~29).
Since entities differ in their means, information ``may be anonymous for some entities, but not for others'' (\S~6), and the assessment starts by asking for whom the data is intended to be anonymous (\S~12).

The guidelines offer two ways to carry out the assessment (\S\S~45--49).
The contextual approach assesses the means of each relevant entity individually, at the risk of overlooking means the assessor is unaware of (\S~47).
The simplified approach disregards differences between entities in access to the given data, additional information and other resources; it overestimates the likelihood of re-identification, which may be more cautious than necessary but ensures that the legal standard is met in cases of doubt (\S~48).
Both approaches apply the same three criteria: No Record Isolation, No Linkage and No Inference (\S\S~52--53).

Of the three criteria, No Inference is the most relevant for aggregate outputs and trained models.
For the given data to relate to an individual, an inference has to be specific and meaningful: it is liable to affect the individual's rights and interests, and it relies on the given data, in the sense that it could not be obtained from general knowledge or from data about the population at large (\S\S~70--71).
Whether an inference relies on the given data can be shown by tracing how it is drawn or, where that is not possible, by comparing two versions of the given data, produced with and without the individual's records (\S~71a--b); Sect.~\ref{sec:n-framework} returns to this comparison.
The criteria have to hold for every individual.
A dataset is anonymous only if the likelihood of re-identification is insignificant for all the individuals in it (\S~36), and a re-identification technique can be successful even if it succeeds against a single individual (\S~87).
The likelihood also changes over time, as techniques improve and more additional information becomes available; if it is no longer insignificant, the data should again be considered personal (\S~35).

A machine learning model is a mathematical function with numerical parameters, called weights, which are adjusted during training until the model reproduces patterns in its training data.
The guidelines treat AI models as data that has been aggregated to represent correlations in the original information, from which inferences about an individual can be drawn by querying or prompting (\S~82).
For a simple model the aggregation is explicit: the few coefficients of a linear regression are computed from summary statistics of the training data, so its weights are aggregate statistics. Large models have billions of weights, which gives them the capacity to retain details of individual training records~\cite{carlini2021extracting}.
A trained model can be handed over as a file of weights, or kept by its provider and offered as a query service, which returns answers without giving access to the weights.
Opinion 28/2024 holds that a model trained on personal data can be considered anonymous only if the likelihood of extracting personal data from it, directly or through queries, is insignificant for any data subject~\cite[\S~43]{edpb2024opinion28}.
It lists attacks that testing could cover, among them membership inference and the reconstruction of training data, and notes that successful testing against known attacks ``can only be evidence for the resistance to those attacks''~\cite[\S~55]{edpb2024opinion28}.

Among these attacks, membership inference is widely studied~\cite{shokri2017membership,carlini2022first}.
The guidelines describe it as inferring that an individual's data was included in the original dataset, which may itself reveal information about the individual or be a first step towards other inferences (\S~79).
As a test, the attacker examines how the model behaves on a record and guesses whether the record was used in training, and the test measures how much better than chance the guesses are.

The standard formal protection against such inference is differential privacy~\cite{dwork2014algorithmic}, a mathematical property of the process that produces a release: it limits how much the output can change when the data of one individual is added or removed, usually by adding random noise.
The limit is set by privacy parameters, and the smaller they are, the stronger the protection and the less precise the results.
The guarantee also depends on the unit of privacy, that is, whether it covers a single record or all the data of an individual, and on which variant of differential privacy is used.

\section{Framework}\label{sec:n-framework}
This section describes the framework used in the open questions and the discussion: a setting in which data is processed and released, the forms a release can take, and the types of evidence that can support an anonymity assessment.

\runin{Secure processing.}
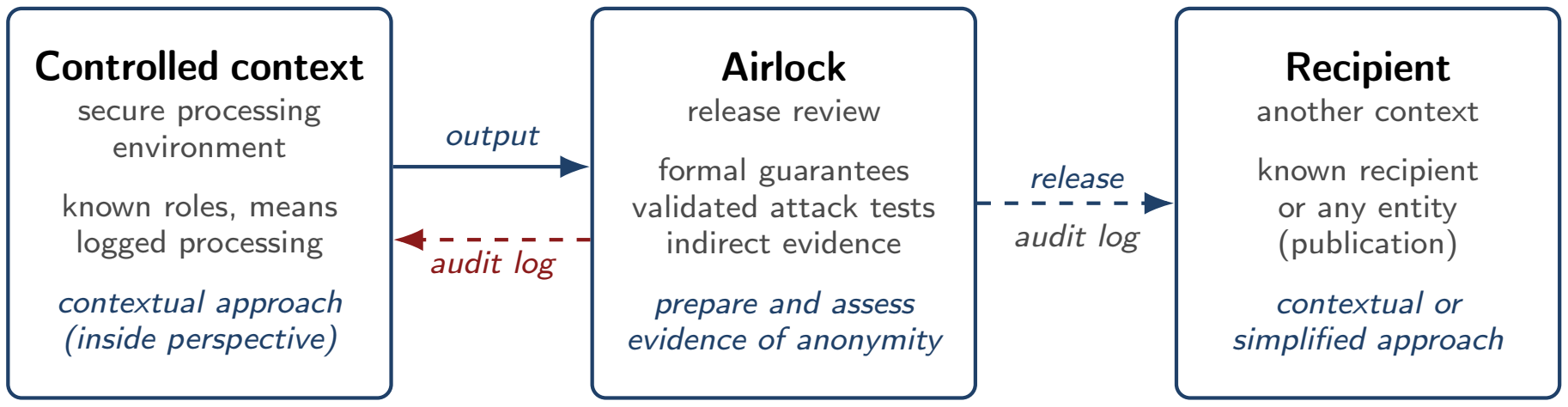
\begin{figure}[t]
\centering
\definecolor{figaccent}{HTML}{1F4068}
\definecolor{figmuted}{HTML}{4A4A4A}
\definecolor{figdanger}{HTML}{8B1A1A}
\begin{tikzpicture}[
  font=\sffamily\footnotesize,
  >=Latex,
  every node/.style={align=center},
  nodebox/.style={draw=figaccent, line width=0.8pt, fill=white,
                  rounded corners=3pt, inner sep=5pt,
                  text width=2.6cm, minimum height=3.0cm, font=\sffamily\scriptsize},
  mut/.style={font=\sffamily\scriptsize, color=figmuted},
  bluelab/.style={font=\sffamily\scriptsize\itshape, color=figaccent, inner sep=1pt},
  redlab/.style={font=\sffamily\scriptsize\itshape, color=figdanger, inner sep=1pt}
]
  \node[nodebox] (spe) at (-4.5,0) {%
    {\footnotesize\bfseries Controlled context}\\[1pt]
    {\scriptsize\color{figmuted}secure processing\\ environment}\\[5pt]
    {\scriptsize\color{figmuted}known roles, means}\\
    {\scriptsize\color{figmuted}logged processing}\\[5pt]
    {\scriptsize\itshape\color{figaccent}contextual approach\\ (inside perspective)}};
  \node[nodebox] (air) at (0,0) {%
    {\footnotesize\bfseries Airlock}\\[1pt]
    {\scriptsize\color{figmuted}release review}\\[5pt]
    {\scriptsize\color{figmuted}formal guarantees}\\
    {\scriptsize\color{figmuted}validated attack tests}\\
    {\scriptsize\color{figmuted}indirect evidence}\\[5pt]
    {\scriptsize\itshape\color{figaccent}prepare and assess evidence of anonymity}};
  \node[nodebox] (out) at (4.5,0) {%
    {\footnotesize\bfseries Recipient}\\[1pt]
    {\scriptsize\color{figmuted}another context}\\[5pt]
    {\scriptsize\color{figmuted}known recipient}\\
    {\scriptsize\color{figmuted}or any entity}\\
    {\scriptsize\color{figmuted}(publication)}\\[5pt]
    {\scriptsize\itshape\color{figaccent}contextual or\\ simplified approach}};
  \draw[->, figaccent, line width=0.8pt] ([yshift=8pt]spe.east) --
        node[bluelab, above=2pt, midway] {output} ([yshift=8pt]air.west);
  \draw[->, figdanger, line width=0.8pt, dashed] ([yshift=-8pt]air.west) --
        node[redlab, below=1pt, midway] {audit log} ([yshift=-8pt]spe.east);
  \draw[->, figaccent, line width=0.8pt, dashed] (air.east) --
        node[bluelab, above=2pt, midway] {release}
        node[mut, font=\sffamily\scriptsize\itshape, below=1pt, midway] {audit log}
        (out.west);
\end{tikzpicture}
\caption{The SPE setting. Inside the controlled context the entities and their means are known, so the contextual approach can be applied from the inside perspective. Each output candidate enters the airlock (solid arrow); both outgoing arrows are conditional on the review: if it passes, the output is released together with its audit log to a recipient, which may be a known recipient (contextual approach) or any entity (simplified approach, in our view); if it fails, the output and log return for revision.}
\label{fig:spe-model}
\end{figure}
A technical platform in which sensitive data can be analysed without leaving the control of its operator is called a secure processing environment (SPE).
Users work with the data inside the environment, and nothing leaves it without review (Fig.~\ref{fig:spe-model}).
We use this setting throughout because it separates the perspective of the entities working inside from that of the recipients of its outputs.
Inside, only users in defined roles have access, their processing is logged and monitored, and their capabilities are restricted: they cannot import external data without permission, and their compute is set by the operator.
Who has access, and with what means, is therefore known, as the contextual approach requires; entities without access need not be assessed (\S~9), and such restrictions on access are among the objective factors to consider (\S~29).
The guidelines treat access to data under appropriate security controls as a means reasonably likely to be used only by authorised users (\S~89), although such measures are not sufficient by themselves (\S~90) and capabilities may change over time (\S~88).
We assume in what follows that the controls are sufficient for the contextual approach to apply inside; whether they are in a given SPE is for the controller to assess.

\runin{Release review.}
The decision on what may leave the environment is taken in an airlock: a review that every output passes before it is released (Fig.~\ref{fig:spe-model}).
An output that passes becomes a release, and one that fails returns for revision.
The review assesses whether the output is anonymous for those who receive it, so the perspective changes from the entities inside to the recipients.
If the output goes to a known recipient that processes it under comparable controls, the assessment can remain contextual, from that recipient's perspective.
If it is published or passed on, any entity may obtain it, and the assessment must consider the means any entity could reasonably use.
For such releases the simplified approach, which a controller may choose in any setting (\S~48), is in our view the natural choice.
The reasoning does not depend on a formal SPE.
It holds for output from any environment with similar controls.
It also applies when data is transferred into such an environment, which is then the recipient.

\begin{table}[t]
\centering
\caption{Types of release.}\label{tab:release}
\scriptsize
\begin{tabular}{@{}L{2.2cm}L{3.2cm}L{5.8cm}@{}}
\hline
\textbf{Release} & \textbf{The recipient receives} & \textbf{Comment}\\
\hline
Records & record-level data set & One record per individual, with identifiers removed and values generalised or suppressed.\\
Aggregates & tables and statistics & Summaries over many individuals, such as counts, averages and regression results.\\
Synthetic data & generated records & Produced by a model trained on the original data. Not treated further here.\\
Model weights & the full model & Can be copied and examined without limit, and cannot be withdrawn once it has been shared.\\
Query service & answers to queries & The provider controls who may query and what is answered, and can withdraw the service.\\
\hline
\end{tabular}
\end{table}

\runin{Types of release.}
Anonymised data can be released in different forms, and each gives a recipient different opportunities for re-identification.
Table~\ref{tab:release} lists the forms we consider, from record-level data to trained models and access to a model through queries.
They differ in what the recipient receives and in what control remains afterwards: records, aggregates and model weights can be copied and cannot be withdrawn, whereas a query service stays under its provider's control, although answers already given remain with those who received them.
Any of the forms may go to a known recipient or to any entity, as set out above.

The forms also differ in how directly they expose individuals.
Record-level data can be attacked through isolation, linkage and inference.
Aggregates summarise many individuals, so the main risk is inference, and several aggregates from the same data can be combined to reveal more than each does alone (\S~77).
A trained model may lie anywhere between an aggregate result and a copy of the original records, depending on how much detail of individual records it retains.
For models as for aggregates the main risk is inference, and membership inference is the most widely used test.

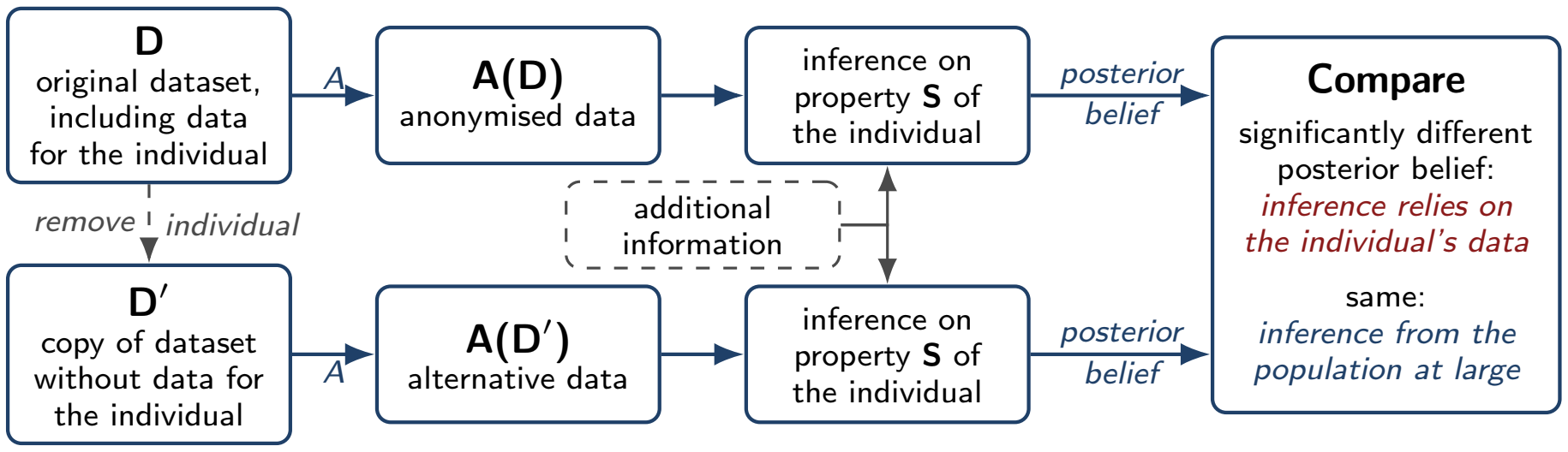
\begin{figure}[t]
\centering
\definecolor{figaccent}{HTML}{1F4068}
\definecolor{figmuted}{HTML}{4A4A4A}
\definecolor{figdanger}{HTML}{8B1A1A}
\begin{tikzpicture}[
  font=\sffamily\scriptsize,
  >=Latex,
  every node/.style={align=center},
  box/.style={draw=figaccent, line width=0.8pt, fill=white, rounded corners=3pt,
              inner sep=4pt, text width=1.9cm, minimum height=1.05cm},
  info/.style={draw=figmuted, line width=0.6pt, dashed, fill=white, rounded corners=3pt,
               inner sep=3pt, text width=1.9cm},
  arr/.style={->, figaccent, line width=0.8pt},
  lab/.style={font=\sffamily\scriptsize\itshape, color=figaccent, inner sep=1pt}
]
  \node[box] (d)  at (0, 1.0)  {{\footnotesize\bfseries D}\\ original dataset, including data for the individual};
  \node[box] (dp) at (0,-1.0)  {{\footnotesize\bfseries D$'$}\\ copy of dataset without data for the individual};
  \draw[->, figmuted, line width=0.6pt, dashed] (d.south) -- node[lab, left=2pt, color=figmuted]{remove} node[lab, right=2pt, color=figmuted]{individual} (dp.north);
  \node[box] (ad)  at (2.85, 1.0) {{\footnotesize\bfseries A(D)}\\ anonymised data};
  \node[box] (adp) at (2.85,-1.0) {{\footnotesize\bfseries A(D$'$)}\\ alternative data};
  \draw[arr] (d.east)  -- node[lab, above]{A} (ad.west);
  \draw[arr] (dp.east) -- node[lab, below]{A} (adp.west);
  \node[box] (i1) at (5.7, 1.0) {inference on property {\bfseries S} of the individual};
  \node[box] (i2) at (5.7,-1.0) {inference on property {\bfseries S} of the individual};
  \node[info] (add) at (4.275, 0) {additional information};
  \draw[arr] (ad.east)  -- (i1.west);
  \draw[arr] (adp.east) -- (i2.west);
  \draw[->, figmuted, line width=0.6pt] (add.east) -| (i1.south);
  \draw[->, figmuted, line width=0.6pt] (add.east) -| (i2.north);
  \node[box, text width=2.4cm, minimum height=2.9cm] (cmp) at (9.55, 0) {%
    {\footnotesize\bfseries Compare}\\[3pt]
    significantly different posterior belief:\\
    {\itshape\color{figdanger} inference relies on the individual's data}\\[4pt]
    same:\\
    {\itshape\color{figaccent} inference from the population at large}};
  \draw[arr] (i1.east) -- node[lab, above]{posterior} node[lab, below]{belief} (cmp.west |- i1.east);
  \draw[arr] (i2.east) -- node[lab, above]{posterior} node[lab, below]{belief} (cmp.west |- i2.east);
\end{tikzpicture}
\caption{The comparison in \S~71b and footnotes~43--44 of the guidelines, in their notation. D$'$ is the original data D without the individual's data; A is the anonymisation process. If an inference on property S works in the same way from A(D) and A(D$'$), it draws on the population at large; a significantly different posterior belief shows that it relies on the individual's data. An attack-based test carries out this comparison, e.g.\ membership inference where S is whether the individual's data was in D.}
\label{fig:inference-test}
\end{figure}

\runin{Types of evidence.}
An anonymity assessment needs evidence that individuals cannot be re-identified to a significant degree, at an airlock or elsewhere.
For the No Inference criterion the guidelines describe a comparison between two versions of the given data: A(D), produced from the original data D, and A(D$'$), produced from the same data without the individual's records (\S~71b, footnotes~43--44; Fig.~\ref{fig:inference-test}).
If an inference about the individual can be drawn only from the first version, it relies on that individual's data.
We call the extent to which an output depends on an individual's data its \emph{leakage}.
Leakage is a property of the output, whether or not anyone exploits it, and should not be confused with a data leak in the sense of a breach.
Evidence about leakage falls into three categories that differ in what they establish and what they cost (Table~\ref{tab:evidence}).
\emph{Formal evidence} is a mathematical guarantee that leakage stays below a stated bound, whatever additional information the recipient holds.
\emph{Empirical evidence} comes from attack-based tests: a successful attack indicates leakage, provided the success is not due to chance or to a biased test, while a failed attack shows only that this attack, under its assumptions, found none.
\emph{Indirect evidence}, such as documentation of how a model was built, informs a judgement about leakage without measuring or bounding it.
\begin{table}[t]
\centering
\caption{Types of evidence.}\label{tab:evidence}
\scriptsize
\begin{tabular}{@{}L{2.2cm}L{3.2cm}L{5.8cm}@{}}
\hline
\textbf{Evidence} & \textbf{Example} & \textbf{Comment}\\
\hline
Formal & differential privacy & Holds for any attacker. Trades utility for protection. Requires a choice of parameters.\\
Empirical & membership inference test, data extraction test & Measures only what the chosen test can detect, which limits its strength as evidence.\\
Indirect & documentation of training, risk indicators & Inexpensive to produce. Supports a judgement but gives no measurement or bound.\\
\hline
\end{tabular}
\end{table}

In our reading, formal and empirical evidence rest on the same comparison: differential privacy bounds how much any inference can differ between A(D) and A(D$'$), and membership inference tests whether an attacker can tell them apart.
Membership inference applies to aggregate results~\cite{homer2008resolving} as well as to trained models, and is our main example.
Every attack-based test assumes some additional information and means for the attacker, which can match the expected attacker or be stronger.
The two approaches enter a test through these assumptions: for an output that stays with known entities, the test can assume their information and means; for an output that may reach any entity, it must assume what any entity could reasonably bring.
For such tests we distinguish the \emph{method} from the \emph{measurement}: the method is the procedure, with its attack and assumptions, which can be validated in advance, independently of any single release, and the measurement is the result of applying it to a given output, reported with its uncertainty. 
When an attack-based test finds no leakage, the output may have little leakage, the attack may be too weak, or the test may fail to detect it for other reasons, for example because it examines too few records or reports only an average.

\section{Open questions}\label{sec:n-points}
This section raises seven open questions, following our reply to the consultation.
Each is given with a short summary, a scenario, the question and our rationale.
Questions~\ref{q:1} and~\ref{q:2} concern empirical evidence, Questions~\ref{q:3} to~\ref{q:5} the setting and form of a release, Question~\ref{q:6} formal evidence and Question~\ref{q:7} indirect evidence.

\runin{Validation of re-identification tests.}
A re-identification attempt that fails is a measurement result, and like any measurement it falls short of proof of anonymity: its strength as evidence depends on what the test can detect.
The documentation of testing (\S~41) should therefore state the detection limit, bias and uncertainty of the test, established by validating the method on cases where the outcome is known.
\begin{scenario}\label{ex:1}
Suppose a test wrongly flags one record in a thousand.
Among 1,000 training records, one flag is then expected by chance even if nothing leaks, and a few more would not stand out.
Such a test can miss leakage that affects several individuals in a thousand.
That level is its detection limit, and a result at chance level shows only that leakage above it is unlikely (see Appendix~B).
\end{scenario}
\begin{question}\label{q:1}
How should the detection limit, bias and uncertainty of a re-identi\-fication test be established and documented when the test is used as evidence of anonymity?
\end{question}
The guidelines judge a re-identification technique by whether its results are close enough, and held with sufficient confidence, to distinguish individuals (\S~87).
By the same reasoning, a test used to show that such techniques fail should state how confident its conclusion is.
Existing standards cover the validation of test methods~\cite{iso17025}, the expression of measurement uncertainty~\cite{jcgm100} and the role of uncertainty when a result is compared with a limit~\cite{jcgm106}.
Other regulated fields require validated methods before the results can support decisions, for example the analytical performance of in vitro diagnostic devices, including their limit of detection~\cite[Annex~I, 9.1(a)]{ivdr2017}, and the independent validation of models in banking~\cite{sr262}.
The guidelines encourage scientifically grounded expertise (\S~84) but do not address the validation of test methods.

\runin{Evidence for individual records.}
The likelihood of re-identification must be insignificant for every individual (\S~36).
Attack-based tests typically gain their statistical power by aggregating over many records and say much less about any single individual.
It is unclear what evidence is expected for the individual.
\begin{scenario}\label{ex:2}
A company tests its model with membership inference before release and finds no significant leakage across the training records.
After release, a data subject withdraws consent and requests erasure.
The pre-release test has little statistical power for this one individual, and once the original training data is deleted, the comparison in \S~71b can no longer be repeated.
\end{scenario}
\begin{question}\label{q:2}
What can be concluded about a single individual from a re-identi\-fication test whose results are aggregated over many records?
\end{question}
Since a technique can be successful even if it succeeds against a single individual (\S~87), evidence is needed at the level of the record, but an estimate for a single record is often too uncertain to show that its risk is insignificant.
The answer may also bear on erasure obligations (cf.\ \S~10), as Scenario~\ref{ex:2} illustrates.

\runin{Secure processing environments with an airlock.}
A secure processing environment (SPE) separates the entities working inside, whose access and means are known and controlled, from the recipients of its outputs, which are reviewed at the airlock before they leave.
\begin{scenario}\label{ex:3}
A company developing an AI model obtains de-identified data from a telecom operator and trains the model inside a secure processing environment.
The company works in defined roles with logged access, cannot import external data and has no access to the telecom operator's additional information.
Under the contextual approach, the data may therefore be anonymous from the company's perspective, although it remains personal data for the telecom operator.
Before the trained model leaves through the airlock, it is tested with attacks on the data still inside.
\end{scenario}
\begin{question}\label{q:3}
How does the contextual approach apply to the entities working inside a secure processing environment, and how should an output be assessed at the airlock from the perspective of its recipients outside?
\end{question}
The guidelines do not mention secure processing environments or the review of outputs that leave them.
In Scenario~\ref{ex:3} the data may be anonymous for the company inside the environment, while the model trained on it goes to recipients whose means may differ.
It is then open when the anonymity of the data carries over to the model (cf.\ \S~21).

\runin{Types of release for AI models.}
The guidelines bring AI models into scope (\S\S~77, 82) and describe the A(D)/A(D$'$) comparison (\S~71b), but do not say whether the given data of a model is its weights or the answers it gives.
Released weights can be copied and attacked with full access, while a model behind a query service releases only its answers and stays under the provider's control.
\begin{scenario}\label{ex:4}
A company has trained a model on its customers' data and wants to offer it to business customers.
It can either ship the model weights for use in the customers' products, or offer access through a query service with authentication, rate limits and answers restricted to labels.
\end{scenario}
\begin{question}\label{q:4}
How does the comparison in \S~71b apply to the release of model weights, compared with access through a query service with restrictions?
\end{question}
Opinion 28/2024 notes that the conclusion of an assessment may differ between a publicly available model and an internal one, and that different levels of testing may be considered~\cite[\S~46]{edpb2024opinion28}.
Recipients of the weights face no limits on who they are or what they do with the model, including inference attacks.
For a service, the provider decides who may send queries, how many and what the answers contain, which restricts access to authorised users (\S~89) and limits what they can obtain from the model.
A service open to the public limits what every user can obtain but not who the users are, which may call for the simplified approach.

\runin{Releases that cannot be withdrawn.}
Re-identification techniques improve over time (\S~35), and the assessment should consider reasonably foreseeable technological developments (\S~29(e)).
Periodic reassessment cannot correct a release that cannot be withdrawn, such as openly published record-level data or model weights.
\begin{scenario}\label{ex:5}
A public agency publishes a record-level dataset under an open licence, and a company publishes model weights in a public repository.
Both have been tested against the attacks known at the time.
A few years later, a stronger attack is published.
Neither release can be withdrawn.
\end{scenario}
\begin{question}\label{q:5}
For a release that cannot be withdrawn, how should reasonably foreseeable developments in re-identification techniques (\S~29(e)) be taken into account at the time of release?
\end{question}
If the likelihood of re-identification later becomes significant, the data should again be considered personal (\S~35).
Attacks can improve quickly: language models, for example, now make linkage of free-text records feasible at scale~\cite{Lermen_DeanonymLLM_2026}.
Because a published release cannot be withdrawn or changed, every mitigation has to be applied before the release.
A model shared with a known partner under contract is in the same position, and the partner may later obtain new additional information, for example through a data breach.
Under the contextual approach, the EDPB recommends adequate safety margins so that the assessment remains robust as the entities' capabilities evolve (\S~88).
The simplified approach disregards differences in means between entities (\S~48), but it is not clear whether it also covers attacks that are not yet known (cf.\ \S\S~29(e), 49).

\runin{Formal guarantees and the simplified approach.}
Differential privacy bounds how much any recipient's posterior belief about an individual can change depending on whether that individual's data was included, whatever additional information the recipient holds.
Because the bound holds for any recipient, it appears to fit the simplified approach.
\begin{scenario}\label{ex:6}
A statistics office publishes counts and averages by region from a sensitive register, produced with a differentially private mechanism, and argues under the simplified approach that the release protects against any recipient.
\end{scenario}
\begin{question}\label{q:6}
Under what conditions can a formal guarantee such as differential privacy support the conclusion that the No Inference criterion is met under the simplified approach?
\end{question}
Differential privacy is not mentioned in the guidelines, although several replies to the 2025 consultation on pseudonymisation raised it~\cite{edpb2025consultation}.
The guarantee still allows inference about the population at large, which the guidelines accept (\S~70).
Several releases from the same data add up (\S~77) and need to be accounted for together.
The meaning of the guarantee depends on the privacy parameters, the unit of privacy and the variant used, and we are not aware of any data protection authority that has set values (cf.~\cite{nist800226}).
Which values are small enough for a posterior belief not to be ``significantly different'' (footnote 44) may depend on the context of the release, even under the simplified approach.

\runin{Indirect evidence.}
Some releases may not need an attack-based test or a formal guarantee.
Indirect evidence, such as the properties of the data and documentation of how a model was built, informs a judgement about the risk without measuring or bounding it.
The guidelines already reason this way in the key factors of \S\S~85 and 86.
\begin{scenario}\label{ex:7}
A hospital releases a simple statistical model: a logistic regression with a handful of coefficients, trained with strong regularisation on several thousand patients.
Its performance on training and test data is almost the same.
The hospital argues that the documentation of the training and these indicators are enough to show that the risk is low, without running attack-based tests.
\end{scenario}
\begin{question}\label{q:7}
For which releases can indirect evidence be sufficient to conclude anonymity without attack-based testing or a formal guarantee?
\end{question}
Indirect evidence is widely used in practice.
Before a model may leave its secure processing environment, Findata, the Finnish authority for the secondary use of health and social data, asks applicants to complete a questionnaire on how the model was built and is to be used, in which attack testing is one question among many~\cite{findata2025anonymous}.
The SACRO-ML tools check model types and settings for disclosure risk before training and run attacks after training~\cite{smith2025sacroml}.
The conditions proposed by Jebreel et al.~\cite{jebreel2026critical} for when an attack poses a realistic threat can serve as a similar check.
Knowing when indirect evidence suffices would help make the depth of assessment proportionate to the risk.

\section{Discussion}\label{sec:n-discussion}
The seven questions are open, and answering several of them requires legal interpretation, which is not the purpose of this paper.
This section examines their technical side and sets out, for each question, what an answer would depend on.
It follows the topics of the paper title: model weights, airlocks and sandboxes.
In each case the assessment comes to rest on the reliability of test results, which leads to the validation of test methods.

\runin{Model weights.}\label{sec:n-weights}
A trained model can leave its provider as weights or through a query service, and Question~\ref{q:4} asks how the assessment differs between the two.
In practice similar tests are used for both, because the best-established membership inference attacks use only the outputs of a model~\cite{carlini2022first,rmia_zarifzadeh2023low}.
The two forms differ in the attacker's means, as set out under Question~\ref{q:4}.
With the weights, an attacker can run any number of queries and inspect the model itself.
Behind a service, queries are limited in number and in what they return, the architecture may be undisclosed, and quantities derived from the weights have to be estimated from the answers. 
A test can mirror these conditions, with the same output format and query budget.
Limits on the number of queries can be circumvented, however, for example by spreading the queries over many accounts, so a test may need a larger budget than a single user has.
Restricting the answers to labels reduces the risk without removing it, because attacks that see only labels can still succeed~\cite{ChoquetteChoo_LabelOnly_2021}.
Withdrawing or changing a service does not take back the answers already given: an attacker can keep them and later apply a stronger attack, or combine them with data that has since become available.
Appendix~A.1 gives the technical detail.

Question~\ref{q:5} concerns releases that cannot be withdrawn, which includes model weights, whether published openly or shared with a known partner; the two are assessed from different perspectives (Sect.~\ref{sec:n-framework}).
The provider of a service can respond when attacks improve: for future queries it can tighten the controls, retrain the model or withdraw it, and periodic reassessment is good practice (\S~35).
A retrained model is a new release that needs its own assessment.
For released weights, the question is then how foreseeable developments of attacks can be allowed for at the time of release.
One way is to add a margin by testing with stronger attacker assumptions than those of the expected attacker, similar to the safety margins that \S~88 recommends under the contextual approach.
Such a margin covers an attacker with more information or resources, but it may not cover a better attack.
The other option is evidence whose bound does not depend on the strength of the attack, which is the subject of Question~\ref{q:6}.

A formal guarantee for model weights can be obtained by training with differential privacy.
Because the guarantee holds whatever the recipient knows or does, it appears to suit a release that cannot be withdrawn; it makes the model less precise, and the conditions under which it can support an anonymity assessment are discussed under Method validation.
The scenario in Question~\ref{q:2}, an erasure request after release, shows why protection applied before release matters.
Unlearning is the removal of an individual's influence from a trained model after the fact.
In its exact form it retrains at least part of the model~\cite{Cao_Unlearning_2015,bourtoule2021unlearning}, which is costly for large models.
Approximate methods can avoid the retraining but are hard to verify, since whether the individual's influence is gone has itself to be established by a test~\cite{xu2023machine,hayes2024inexact}.
Removing some records can also leave others more exposed~\cite{carlini2022privacyonion}.
It is therefore an open question whether preventive measures at training time, such as noise addition and regularisation, combined with a test before release, are preferable to unlearning followed by a new test~\cite{LeakPro24}.
Whether either would satisfy an erasure obligation is a legal question, which the guidelines leave aside (\S~10).
For released weights, the assessment therefore depends on evidence gathered before the release.

\runin{Airlocks.}\label{sec:n-airlocks}
Some models are trained in a controlled environment, where those inside have access to the training data and those outside receive only the model, as in Scenario~\ref{ex:3}.
The setting may become more common as access to compute for training grows, for example through the AI factories set up under the EuroHPC Joint Undertaking~\cite{eurohpc2024aifactories}.
Where such training uses another organisation's sensitive data, the environment may be provided by the facility or by another party, and the model has to pass a release review before it leaves.
Established procedures exist for checking aggregate outputs~\cite{griffiths2019sdc}, and recent work extends them to trained models~\cite{jefferson2022graimatter,smith2025sacroml}.
Testing before release, while the original data is still available and it is known which records were used in training, gives the best opportunity to gather evidence.

The assumptions of such a test follow from who is to receive the output, as set out in Sect.~\ref{sec:n-framework}.
The setting also allows a mitigation to be assessed, for example noise added to a statistic or a model trained with differential privacy, since the same test can be run before and after the mitigation is applied.
The test itself is processing of the original data.
When it is carried out inside the environment before release, the test is subject to the same controls as the processing it assesses.
The guidelines presume the legal basis of anonymisation to coincide with that of the preceding processing when it is part of the same processing activity and pursues the same purposes (\S~38).
In our reading, this extends to an attack test carried out before release, since the guidelines count the testing of supposedly anonymous datasets as part of the anonymisation process (\S~41; cf.~\cite{curelariu2024lawfulness}).
Deleting the source data after processing is good practice in secure processing, but once it is deleted the comparison in \S~71b can no longer be repeated (the situation in the erasure scenario of Question~\ref{q:2}), and the guidelines then point to a rebuttable presumption of non-reliance (\S~71c).
An archived test report still allows the assessment to be reviewed, but not repeated.
The anonymity of data inside the environment can carry over to a model trained on it if the release review establishes that the recipient has no additional information or means beyond those of the entities inside; otherwise the model needs its own assessment (Question~\ref{q:3}; cf.\ \S~21).
Appendix~A.2 gives the setting in full and extends it to federated learning, where the updates exchanged between parties amount to repeated releases of weights.

\runin{Sandboxes.}\label{sec:n-sandboxes}
A regulatory sandbox is a supervised framework in which a competent authority guides a project that develops and tests an innovation~\cite{gumbo2025sandbox,baldini2024sandboxes}.
Examples are the AI regulatory sandboxes of the AI Act~\cite{EU_AI_Act_2024,Burden_Stenberg_2025} and sandboxes run by national data protection authorities~\cite{IMY_FedML_2023,SARDIN24}.
Sandboxes vary: some centre on guidance on how rules apply, and others also include technical testing that produces evidence for the authority~\cite{Burden_Stenberg_2025,buscemi2026configurator}.
Our interest is in a narrower group, in which personal data, pseudonymised or not, may be processed in a controlled environment under a legal ground of its own.
The AI Act provides for this under certain conditions, among them a substantial public interest (Art.~59), and the European Health Data Space does so in its secure processing environments, which are not sandboxes but share this feature~\cite{Burden_Stenberg_2025}.
In the latter, under Art.~73(2), the health data access body reviews each download request so that only non-personal data leaves the environment~\cite{EHDS_2025}.

These settings differ from that of Sect.~\ref{sec:n-framework}, where the data inside may itself be anonymous for those who work with it.
Here controls are still required, but anonymity is not required inside, and we assume that what leaves must be anonymous for those outside.
The release review is then the single point at which anonymity is assessed, and what remains of Question~\ref{q:3} is how an output is assessed for those outside.

With an authority present, the conditions of a test can be agreed before it is run.
A sandbox project follows an agreed plan~\cite{Burden_Stenberg_2025}, and it is a question whether a test protocol, with its attacks, metrics and thresholds, could be made part of that plan and reviewed in advance.
This bears on who chooses the tests: a provider who selects them has a conflict of interest, and in the engagements described by Buscemi et al.\ the authority retains the power to mandate a minimum set~\cite{buscemi2026configurator}.
Running a test is itself processing of personal data.
Art.~59 of the AI Act allows processing in the sandbox for developing, training and testing an AI system; whether a test of the anonymity of its outputs falls under this is a legal question that we leave open.
The outcomes of a sandbox include documentation and test results~\cite{Burden_Stenberg_2025}, and the authority needs evidence it can rely on~\cite{buscemi2026configurator}; how far a test result can be relied on is the subject of Question~\ref{q:1}.
Appendix~A.3 gives an overview of these regimes and their provisions.

\runin{Method validation.}\label{sec:n-validation}
If a test result is treated as a measurement, as in Question~\ref{q:1}, three plain questions follow: how much leakage could the test have detected, does it over- or underestimate, and how certain is the result?
In measurement terms, these are the detection limit, bias and uncertainty of the test.
The detection limit depends on the strength of the attack: a weak attack fails even where there is leakage, so a test built on one can overstate protection (cf.~\cite[\S~55]{edpb2024opinion28}), as Hanisch et al.\ show for biometric anonymisation~\cite{hanisch2024false}.
Bias arises when the test differs systematically from the release, for example when the tested records are more typical than the individuals most at risk~\cite{aerni2024misleading}.
The uncertainty depends on how many records are tested and how often the test is repeated, and it is largest at the low false positive rates that matter most~\cite{carlini2022first}.
Beyond the general standards named under Question~\ref{q:1}, several standards address de-identification~\cite{iso20889,iso27559}, and two NIST publications discuss testing, including motivated intruder tests and empirical audits of differential privacy~\cite{nist800188,nist800226}. The first asks agencies to document that the tests used were validated for their purpose, but neither says how such validation is done or what it should establish.

A test method should be validated in advance for its intended use, a class of releases and an attacker, in reference cases where the leakage is known; this is how its detection limit, bias and uncertainty are established.
Anonymeter, for example, validates its measurements on synthetic data by inserting known leakage and checking that the measured risk follows it~\cite{giomi2023anonymeter}, and canary-based auditing of differentially private training inserts records with known membership to measure attack error rates and a lower bound on the privacy parameters~\cite{nasr2023tight,steinke2023one}.
A validation also shows how a result depends on the resources of the test.
Many membership inference attacks~\cite{carlini2022first,rmia_zarifzadeh2023low} compare the model with reference models trained for the purpose, and our own study across twelve test settings examines how the success of established attacks depends on the number of such models, and which attack is reliable at a given number~\cite{brannvall2026bavaria}.
The assessment of a particular release is a separate step.
It applies the validated method and includes checks that the conditions of the validation hold, for example that the tested records represent the individuals in the data and that the attack has converged.

Question~\ref{q:2} asks what such a test shows about a single individual.
The risk may be concentrated in a few individuals, often outliers in the data, whom an average over all records hides.
A test can therefore report results for the most vulnerable records, for example the true positive rate at a low false positive rate~\cite{carlini2022first}, which is the share of training records identified when very few records are flagged wrongly.
Repeated experiments can add evidence for individual records, but their interpretation depends on how well the experiments represent the release.
When few or no successes are observed, the result gives an upper confidence bound on the success of the tested attack, so a risk above zero remains possible.

A differential privacy guarantee is proved for a mechanism and holds whatever attack is tried, provided that the mechanism is implemented correctly.
This can be checked by privacy auditing with canaries, as described above~\cite{steinke2023one}.
An audit is itself a test, and a weak one can miss a faulty implementation, so it needs validation in the same way as the tests of Question~\ref{q:1}.
A validated audit could be one of the conditions that Question~\ref{q:6} asks for.

The three types of evidence complement each other.
A formal guarantee gives an upper limit on leakage and a validated attack-based test a lower one, so the leakage of the release lies between the two.
For large machine learning models, a strict limit often makes the predictions too imprecise, so a looser limit is chosen.
Such a limit can still be conservative under realistic assumptions about the release, with the leakage well below it.
A test then indicates what an attack achieves in practice, and for releases without a guarantee it is the main empirical evidence.
Indirect evidence is less firm, but it costs little and is sometimes all that is available.
The three can be combined by assessing a release in stages, as the SACRO-ML tools do (Question~\ref{q:7}).
The first stage examines indirect evidence, such as documentation and risk indicators.
This can screen out low-risk releases, as the hospital argues in Scenario~\ref{ex:7}, and where the risk is higher it gives context for the empirical tests that follow.
The depth of testing can then be proportionate to the risk and intended use of the release (cf.~\cite[\S~46]{edpb2024opinion28}).
Where an attack test is run, a validated attack from the current state of the art makes the results reproducible and comparable. As with model validation in finance~\cite{sr262}, the test should be carried out with sufficient independence from the release decision.
Validated tests could also help to establish for which releases indirect evidence alone is enough, since they can show on reference cases how well the risk indicators predict what a test finds.
Appendix~C gives an example of such a staged assessment.

The guidelines ask for documentation of the anonymisation process, including the testing (\S~41), and Opinion 28/2024 expects reports on how a model has been tested, by whom, when, how and to what extent, together with the results~\cite[\S~58]{edpb2024opinion28}.
Neither says what a report needs to contain for its results to be assessed.
Following the discussion above, a report would give the measurement with its uncertainty and describe the method used: the form of release and the attacker assumed, the protocol set in advance, and a reference to the validation of the method.
It would include results for the most vulnerable records, and it would name the conditions under which the release should be assessed again, such as new attacks or a change to the release (\S~35).
The items are listed in Appendix~C.

\section{Conclusions}\label{sec:n-conclusion}
The guidelines make testing part of an anonymity assessment.
We have examined the technical side of seven open questions for model weights, airlocks and regulatory sandboxes, and in each setting the assessment depends on how far a test result can be relied on.

What we recommend concerns measurement practice.
Test methods should be validated before their results are used as evidence, a release should come with the measurement, its uncertainty and a description of the method, and the test should be carried out independently of the release decision.
Other matters cannot be settled by measurement and are questions for the legal side. Among them are what the documentation of testing (\S~41) is required to contain, whether a test protocol can be reviewed in advance, which values of the privacy parameters are small enough, how small the risk for a single individual must be to count as insignificant, and whether the legal ground for processing in a sandbox covers tests of anonymity.

\newpage
\bibliographystyle{splncs04}
\bibliography{refs}

\section*{Acknowledgements}
We thank colleagues and collaborators in the Swedish AI and public-sector communities for discussions on the privacy testing of machine learning models and on regulatory sandboxes, which have informed this paper.
The views expressed are those of the authors.

\section*{Use of AI tools}
The authors used Claude (Anthropic; models Claude Opus 5.5 and Claude Fable 5.2) to search and summarise literature, to check citations against their sources, to draft and edit text from the authors' outlines and instructions, including the appendices, to write \LaTeX{} code for the tables and diagrams, and to review the manuscript.
No AI-generated images are used.
The authors have reviewed and verified the content and take full responsibility for it; the positions and arguments are their own and come from their reply to the consultation~\cite{rise2026reply}.

\appendix
\section{Extended discussion and related work}\label{app:discussion}
The appendices add detail and related work to the main text.
Appendix~\ref{app:discussion} opens with notes on the scope and limitations of the paper and on the two forms of release that the main text leaves aside, and then follows the four parts of Sect.~\ref{sec:n-discussion} (A.1 to A.4);
Appendix~\ref{sec:app-detection} works through the example behind Scenario~\ref{ex:1};
Appendix~\ref{app:staged} describes a staged assessment and a test report;
and Appendix~\ref{app:dp} gives more detail on differential privacy.

A reader who has to decide on a release can start with Appendix~\ref{app:staged}, which describes many practical and organisational aspects.
Appendix~\ref{sec:app-detection} and Appendix~\ref{app:dp} both provide technical background.
A.2 and A.3 concern those who run secure processing environments with disclosure controls, or regulatory sandboxes.

\runin{Scope.}
The paper concentrates on the No Inference criterion because it matters most for aggregates and trained models.
Isolation and linkage mainly concern record-level data, for which risk-based assessment frameworks exist~\cite{iso27559}, although the guidelines note that inference by linkage also applies to aggregate data (\S~81).
Membership inference is our main example because the guidelines name it (\S~79) and it is widely studied in the testing literature~\cite{shokri2017membership,carlini2022first}.

Several questions from the wider debate are left aside.
We do not discuss whether the risk of a release should be weighed against its utility, a point raised in many industry replies to the 2025 consultation on pseudonymisation; the guidelines do not weigh the two, and count the value of the re-identified data to an entity only as one factor in the means reasonably likely to be used (\S~31).
We take the contextual and simplified approaches as given and do not return to the earlier debate on relative and absolute identifiability.
We also take from the guidelines which means have to be considered, including unlawful ones (\S\S~32--33); in a test this becomes an assumption about the attacker, which has to be stated.
Data subject rights come up only where a test bears on them, as with erasure in Scenario~\ref{ex:2}.
Anonymisation techniques are not compared~\cite{iso20889}, since the paper concerns the testing of the outcome.

\runin{Limitations.}
The paper is a position paper.
Its scenarios are illustrative and do not cover every form of release, and we have not evaluated test methods empirically in this work.
Where the appendices read the guidelines or other legal texts, the readings are ours and are marked as such; they have not been examined against case law.
A useful next step would be an open library of reference cases with known leakage, on which test methods can be validated and compared (A.4, Who validates).

\runin{Records and aggregates.}
Sect.~\ref{sec:n-discussion} discusses model weights and query services, and leaves aside records and aggregates, two of the other forms of release in Table~\ref{tab:release}.
Record-level data can be assessed with the contextual approach where the recipients and their means are known, as inside an SPE.
Open release is more difficult, because sparse records with many attributes can be re-identified by matching them against public data~\cite{Narayanan_Netflix_2008}, and language models now make the same kind of linkage feasible for free text at scale~\cite{Lermen_DeanonymLLM_2026}.
Where anonymisation cannot be made reliable, protection has to come from access control.
Record-level anonymisation is also used as a step before analysis or training, including the training of a generator of synthetic data, which this paper does not cover.

Aggregates such as counts, averages and charts are the usual output of statistical analysis, and disclosure control for them is well established~\cite{griffiths2019sdc}.
The guidelines note that techniques exist to reveal information about individuals from aggregates (\S~17), especially when several are produced from the same data (\S~77).

\subsection{Model weights}\label{app:weights}
Sect.~\ref{sec:n-discussion} compares the release of weights with a query service (Question~\ref{q:4}) and asks how a release that cannot be withdrawn can allow for future attacks (Question~\ref{q:5}).
This part adds detail in the same order: the attacks and the tests built on them, query services, attacks that appear after a release, and unlearning.
Formal guarantees, which Sect.~\ref{sec:n-discussion} mentions for model weights and takes up under Method validation, are treated in Appendix~\ref{app:dp}.

\runin{Attacks on trained models.}
Several kinds of attack have been developed.
Membership inference estimates from a model's behaviour whether a record was in the training data~\cite{shokri2017membership,carlini2022first}.
Extraction recovers training data from a model~\cite{carlini2021extracting}, model inversion reconstructs features of the training data~\cite{Fredrikson_ModelInversion_2015}, and gradient inversion reconstructs inputs from the updates exchanged in federated learning~\cite{Zhu_DLG_2019}.
The leading membership inference attacks are the likelihood ratio attack (LiRA)~\cite{carlini2022first} and the robust membership inference attack (RMIA)~\cite{rmia_zarifzadeh2023low}.
Both compare the model under test with reference models, which are models trained on similar data.
In the setup of Carlini et al., each reference model is trained on a random half of the data, so that every record is in the training data of about half of them and absent from the rest.

\runin{Models as personal data.}
The legal debate on AI models has asked whether a model memorises personal data.
Veale, Binns and Edwards argued early that model inversion and membership inference show that training data can in some cases be recovered from a model, so that models vulnerable to such attacks could be classified as personal data~\cite{veale2018remember}.
Nolte, Finck and Meding argue that a large language model from which memorised personal data can be extracted by means reasonably likely to be used is itself personal data, with consequences for every use of the model~\cite{nolte2025llms}.
In a discussion paper, the Hamburg data protection authority takes a different position for large language models: they do not store personal data, so data subject rights cannot relate to the model itself, while the GDPR applies to training and to the inputs and outputs of a system~\cite{hamburg2024llm}.
Arga e Lima accepts that models memorise personal data and argues for a contextual assessment, in which a model is anonymous for the parties that lack reasonable means to extract the data~\cite{arga2025memorisation}.
Opinion 28/2024 turns the question into one of likelihood, to be supported by testing~\cite[\S\S~43, 55]{edpb2024opinion28}.

Whether a model memorises is a question about what the weights contain, and for large models it cannot in practice be answered by inspection.
A membership inference test answers a different question: whether the model's behaviour depends on an individual's data to a degree that an attacker can detect.
The Hamburg paper makes a similar point: such attacks say nothing about storage in the model, only about the probability that data was used in training~\cite{hamburg2024llm}.
The comparison in \S~71b asks this second question, and a test estimates the answer.

\runin{Attacks as means.}
In the terms of the guidelines, an attack is a means.
Means are understood broadly, up to ``running incredibly complex analyses and inferences'' (\S~28).
In Example~7, a technique that a business partner could not apply itself, but could hire a third party to apply, still counts as a means reasonably likely to be used.
Relevant factors include how vulnerable the data is to the different means (\S~29(a)) and the available technology with its reasonably foreseeable developments (\S~29(e)).
In our reading, a published attack that a recipient could run, or have someone run, therefore has to be considered in the assessment, even if the recipient lacks the skill to run it.

\runin{Tests for weights and for query services.}
With released weights the attacker has the model itself, yet the best-established attacks use only its outputs~\cite{carlini2022first,rmia_zarifzadeh2023low}.
Tests that exploit the weights directly are less established.
Under an idealised model of training, the optimal attack uses the model only through its loss on the target record, that is, how well the model fits it, compared with the loss typical of models trained without the record~\cite{sablayrolles2019whitebox}.
In that setting access to the weights adds nothing to an attack that already sees the loss; in practice this holds approximately, which is one reason why similar tests are used for both forms.
This result concerns membership inference.
Attacks of other kinds may gain more from access to the weights: in some cases a significant part of the training data has been reconstructed from the parameters of a trained classifier~\cite{haim2022reconstructing}.
A test built on membership inference therefore does not cover everything that a release of weights exposes.

The provider of a query service can authenticate users, limit the number and type of queries, monitor use and restrict what the answers contain.
These controls restrict the means of a user, which is how the guidelines treat security controls: access becomes a means reasonably likely to be used only for authorised users (\S~89).
They also change what is released, since the recipient receives answers instead of the model.
In our reading the assessment covers both: the answers, as the data that is released, and the controls on how answers are obtained.
A test can mirror the service, with the same output format and query budget, but the restrictions do not remove the risk.
If the service returns only labels, each answer carries less information, yet attacks that see only labels can still succeed with more queries~\cite{ChoquetteChoo_LabelOnly_2021}.
A limit on queries applies to an account, and an attacker can open many accounts and spread the queries over them.
This has been reported for model extraction, in which the answers of a service are used to train a model that imitates it: one provider describes campaigns that together used about 24,000 fraudulent accounts for over 16 million exchanges~\cite{anthropic2026distillation}.
A test that assumes the query budget of a single user can therefore understate what an attacker obtains (Sect.~\ref{sec:n-discussion}).
The same holds for data query systems, which the guidelines mention among the sources from which information about individuals can be inferred, often by combining several results produced from the same original data (\S~77): each answer is a release, and the provider's controls decide how many answers an attacker can collect.

\runin{Attacks not yet known.}
Question~\ref{q:5} asks how a release that cannot be withdrawn can allow for attacks that appear later.
The guidelines do not state whether the simplified approach covers attacks that are not yet known.
Our reading of the text is that it does not, though the point is open.
Under the simplified approach the controller ignores differences between entities ``regarding access to the given data, to additional information or to other resources'' (\S~48) and assumes ``that if means reasonably likely to be used exist then at least someone will be able to use them'' (\S\S~100, 103).
Both formulations concern who can use a means, and do not address which means exist or will exist.

On this reading, a method that is known today counts as available to someone, provided it is reasonably likely to be used.
A method that is not yet known would come in through the ``reasonably foreseeable technological developments'' that the guidelines list among the factors to consider (\S~29(e)), under either approach.
An attack that nobody could foresee would then fall outside both.
For data that can be reassessed, the guidelines provide for this: if the likelihood is no longer insignificant, the data should again be considered personal (\S~35).
For a release that cannot be withdrawn the same would apply, but the copies already made stay where they are, which is the concern behind Question~\ref{q:5}.

A broader reading is possible: in an example of how the two approaches combine, the guidelines describe the simplified approach as asking whether re-identification is ``even possible in theory'' (\S~49), which could be taken to include any attack, present or future.
Against this, the assumption in \S\S~100 and~103 keeps the qualification ``reasonably likely to be used''.
Which reading is right is a legal question that we leave open, as in Question~\ref{q:5}.

\runin{Assumptions about the attacker.}
A test does not always assume the attacker that is expected in practice.
What a real attacker knows may be uncertain, and it is then easier to say what the attacker cannot exceed than what the attacker actually has, so a test can assume that upper limit.
A strong assumption can also be the practical choice: if the attacker is taken to know all the training data except the target record, the test is simple to specify and to repeat.
Sect.~\ref{sec:n-discussion} suggests such stronger assumptions as a margin, similar to the safety margins that \S~88 recommends under the contextual approach.
Since \S~88 concerns the capabilities of entities changing over time, reading it as support for stronger test assumptions is our own step.

A test that assumes a strong attacker errs on the side of caution, because an attacker with less knowledge cannot do better than the best attack available to one with more.
If the test finds no leakage, the result also covers attackers who know less, provided that the attack that was run is close to the best one under the assumptions.
If it finds leakage, the risk from the expected attacker may be lower.

The margin covers an attacker with more knowledge, but not a better method of attack.
Once the attacker's knowledge is fixed, the best test is known in principle: it is a likelihood ratio test~\cite{carlini2022first}.
For realistic models it can only be approximated, how close an approximation comes is an empirical matter, and a test is also limited by what the tester can spend, for example on reference models~\cite{brannvall2026bavaria}.
A result therefore has to be read together with the attacker that was assumed, which the test report should state (Sect.~\ref{sec:n-discussion}).
A formal guarantee, in contrast, holds for any attack (Appendix~\ref{app:dp}).

\runin{Unlearning.}
Sect.~\ref{sec:n-discussion} uses the erasure request of Scenario~\ref{ex:2} to show why protection has to be applied before release, and leaves open whether preventive measures or unlearning are preferable.
The literature raises three concerns about unlearning: its cost, how it is verified, and its effect on other individuals.

Exact unlearning retrains the model without the record.
It can be made cheaper in two ways: by writing the learning algorithm so that the model depends on the data only through a few sums, from which a record's contribution can be subtracted~\cite{Cao_Unlearning_2015}, or by training the model in separate parts, so that only the part that saw the record is retrained~\cite{bourtoule2021unlearning}.
The first does not extend easily to deep neural networks, and the second costs accuracy on complex tasks~\cite{bourtoule2021unlearning}.

Approximate unlearning adjusts the trained model instead, so that it behaves as if the record had not been used.
Whether the record's influence is gone then has to be shown by a test, for example a membership inference test on the removed record.
Hayes et al.\ find that the tests commonly used for this overestimate the protection: an attacker who tailors the attack to each record does considerably better than one who uses the same attack for all records~\cite{hayes2024inexact}.
This is the problem of Question~\ref{q:2}, since the claim concerns one record and a test averaged over many says little about it.

The third concern arises even when a record is removed completely.
Carlini et al.\ train models, identify the tenth of the records most exposed to a membership inference attack, remove them from the data, and train new models from scratch on what remains~\cite{carlini2022privacyonion}.
In the new models, records that were safe before are exposed to the same attack, which the authors call the onion effect.
The explanation they propose is that a record's exposure depends on the other records in the data: once the most unusual records are gone, others become the unusual ones.
Removing records can therefore increase the exposure of others, and a test result that looks safe can cease to hold when the data changes.
Hayes et al.\ observe an increase for approximate unlearning as well: with two recent methods, leakage increased for more than half of the records that remained~\cite{hayes2024inexact}.

An alternative is to protect records during training.
One proposal identifies the most exposed records with a membership inference test, reduces their weight in training and adds noise to all training steps, and reports a clear reduction of the risk, measured with the same attack on one benchmark, without a significant loss of accuracy~\cite{LeakPro24}.
The report also considers the onion effect.
With noise and reweighting, the risk scores are lower both for the most exposed records and for the rest, with no visible pattern of risk moving to other records, although this was not examined in detail~\cite{LeakPro24}.
Carlini et al.\ suggest more generally that defences which do not train with rigorous privacy guarantees are unlikely to be effective, although their experiments concern the removal of records~\cite{carlini2022privacyonion}.
Which approach gives an individual better protection, and at what cost to the model, is open.
Whether either would satisfy an erasure obligation is a legal question; the guidelines give no guidance on provisions that mandate erasure (\S~10).

\subsection{Airlocks}\label{app:airlocks}
Sect.~\ref{sec:n-discussion} describes training inside a controlled environment with a review before release, and asks when the anonymity of data inside carries over to a model (Question~\ref{q:3}).
This part adds detail on the practice of output checking that the review comes from, on the environment and what it allows a test to do, on what follows for the insiders and for a model that leaves, and on decentralised environments.

\runin{Output checking.}
The review of outputs before release is established practice for statistical results, under the name of statistical disclosure control.
The Five Safes framework breaks a decision on data access into five dimensions (safe projects, people, data, settings and outputs), which are assessed separately and judged together.
``Safe outputs'' covers the risk that remains in released results, and it is most easily controlled where a facility can check every result before release~\cite{desai2016five}.
A practitioner handbook gives rules of thumb for each kind of output, such as a minimum number of observations behind a figure, to be applied with judgement by trained checkers who are independent of the analyst~\cite{griffiths2019sdc}.
It covers tables, graphs and regression results, and does not address machine learning models.

For trained models the practice is recent.
Mansouri-Benssassi et al.\ interviewed twenty such environments and found no guidelines, mature processes or tools for assessing models~\cite{mansouri2021tre}.
Ritchie et al.\ explain why models do not fit the established checks: a model is too large to inspect, it may include training data, and its risk has to be judged from calculations for which no standard of acceptable risk is agreed~\cite{ritchie2023tre}.
The GRAIMATTER recommendations ask that staff of the environment check a model before it leaves.
The checks include simulated membership and attribute inference attacks on data held back from the researcher, since the vulnerability of a model cannot be predicted in advance, and the recommendations ask that average measures are not relied on alone~\cite{jefferson2022graimatter}.
The SACRO-ML package implements such checks: before training it warns about settings that are likely to lead to disclosure, and after training it runs membership and attribute inference attacks, LiRA among them~\cite{smith2025sacroml}.

Two frameworks from the United Kingdom treat anonymity as dependent on context, as the contextual approach does.
The Anonymisation Decision-Making Framework holds that ``whether data are or are not anonymised is not a property of the data, but determined by the relationship between the data and the context(s) in which they are held''~\cite{elliot2020adf}.
It describes penetration tests by ``friendly'' intruders.
The guidance of the UK Information Commissioner requires a ``motivated intruder'' test, with an intruder who is reasonably competent and has access to public resources~\cite{ico2025anonymisation}.

These sources take up the reliability of such tests to different degrees.
The Anonymisation Decision-Making Framework gives reasons for caution in reading the result of a penetration test: it is a snapshot, it depends on the skill and time of the tester, and it is specific to the dataset.
It does not set out how the reliability of a test is established.
The ICO guidance does not address how a motivated intruder test is validated.
GRAIMATTER asks that simulated attacks ideally come with confidence intervals, and the paper that describes SACRO-ML does not report them.
Question~\ref{q:1} starts from this point.

\runin{The environment.}
Sect.~\ref{sec:n-framework} describes a secure processing environment by its controls: defined roles, logged and monitored processing, no import of external data without permission, and compute set by the operator.
Scenario~\ref{ex:3} places a company in such an environment because training needs more compute than the company has.
The AI factories set up under the EuroHPC Joint Undertaking are one source of such compute.
The Regulation defines an AI factory as an entity that provides ``an AI supercomputing service infrastructure'' with associated services, among them support for the training, testing, evaluation and validation of AI models~\cite{eurohpc2024aifactories}.
It does not mention personal data or secure environments.
Announcements by the Joint Undertaking say that individual factories will support sensitive data; one is to put ``special emphasis on AI applications involving highly sensitive data such as health information''~\cite{eurohpc2025sixfactories}.
The implementing regulation on AI regulatory sandboxes names AI Factories among the facilities with which the authorities are to cooperate, and provides for ``regulatory supervision of specific projects within the AI regulatory sandboxes that may be carried out in those facilities'' (Art.~4(5))~\cite{eu2026_2487}, so a sandbox project may take place in such a facility (A.3).
Whether a given facility offers the controls that Sect.~\ref{sec:n-framework} assumes has to be established for that facility.

The parties in Scenario~\ref{ex:3} stand in different positions.
The telecom operator holds the original data and the additional information that links it to customers, and for the operator the data remains personal.
The company that develops the model works inside the environment without that additional information.
It decides the purposes and means of its own processing, so whether the data is personal for the company is assessed from the company's perspective.
The guidelines give a similar case in Example~4, where a hospital sends excerpts from patient records to an independent research institute that does not act on its instructions.
An entity that processes data on behalf of another is treated differently: for it the question is assessed from the perspective of the controlling entity (\S~15).
In our reading this matters for whoever provides the environment, if that party processes the data on behalf of one of the others.
The assumption in Sect.~\ref{sec:n-framework} that the contextual approach applies inside concerns entities acting on their own account.
Whether the telecom operator may share the data in the first place is a legal question that we leave aside.

The contextual approach can be used in other settings as well.
In a secure processing environment its conditions can be shown from the controls, and the further a setting is from such controls, the harder the assumptions about the entities and their means are to support with evidence.

\runin{Testing inside the environment.}
Sect.~\ref{sec:n-discussion} says that testing before release, while the original data is available and it is known which records were used in training, gives the best opportunity to gather evidence.
One reason is the comparison group.
A membership inference test needs records that were not used in training and that resemble the training records in every other respect.
Evaluations of large models have been shown to be flawed on this point.
Meeus et al.\ find that collections of members and non-members assembled after a model's release differ systematically~\cite{meeus2025sok}, and Das et al.\ show for eight published collections that an attack which does not look at the model does better than the attacks being evaluated~\cite{das2025blind}.
Both recommend a random split made before training.
Inside the environment this can be arranged: a random part of the data is set aside before training and serves as the non-members.
GRAIMATTER recommends that the environment keeps such data and never gives it to the research team~\cite{jefferson2022graimatter}.
Testing with records that were set aside is a practical way to approximate the comparison in \S~71b, since it compares the records in bulk instead of producing a model with each individual's data removed separately.
The test needs the original data; once it is deleted, the guidelines point to the presumption in \S~71c (Sect.~\ref{sec:n-discussion}).

\runin{The time an entity has access.}
An environment that controls who has access also controls for how long.
The period is set by an access agreement or a data permit, and the environment can end access when the period is over.
A data-sharing agreement with a deletion clause relies instead on the recipient to delete.
While access lasts, the provider of the environment keeps control, and the assessment can be repeated as the guidelines recommend (\S~35).

In our reading this bears on Question~\ref{q:5} for the entities inside.
Among the factors for the means reasonably likely to be used, the guidelines list ``the available technology at the time of the processing, as well as reasonably foreseeable technological developments'' (\S~29(e)).
For an entity whose access ends on a known date, the developments to foresee are those of the period of access.
For a release that cannot be withdrawn there is no such date.

The reading covers only the access itself, and four things fall outside it.
People may remember what they have learned inside the environment, and can combine it with information they acquire after their access ends.
Outputs that passed the release review remain with their recipients; they are assessed for those recipients, and for outputs that cannot be withdrawn the concern of Question~\ref{q:5} remains.
Data kept for audit and reproducibility is not deleted when access ends; in our reading, where the controller holds it, it remains personal data for the controller and subject to storage limitation (Art.~5(1)(e) GDPR).
Finally, the reading assumes that no copy leaves the environment by another route: the guidelines ask whether restrictions on access are ``effective and permanent'' (\S~29(b)), and a security incident may lead to a reassessment (\S~42).

\runin{Carry-over of anonymity.}
Sect.~\ref{sec:n-discussion} says that the anonymity of data inside the environment can carry over to a model trained on it ``if the release review establishes that the recipient has no additional information or means beyond those of the entities inside; otherwise the model needs its own assessment''.
The statement is ours and rests on a technical argument that gives a sufficient condition; whether meeting the condition is enough in law is a question we leave open.

Suppose that the data is anonymous for the entities inside, who hold the training data, the model trained on it and their own additional information.
A model computed from the data carries no more information about the individuals than the data does~\cite{cover2006elements}.
If everything the recipient knows is also known inside, whatever the recipient can infer from the model could therefore be inferred inside as well.
The same holds for means other than information, such as compute, skill, time and legal access, as long as the recipient has none beyond those available inside.
Under that condition the model does not help the recipient to identify anyone whom the entities inside cannot identify.
What carries over is the anonymity that the data has for the entities inside, whose means are limited by the controls of the environment.
The condition can therefore be met by a recipient who is limited in the same way, and an organisational measure on the recipient's side counts to the extent that it limits the recipient's information and means.

The condition is demanding, for four reasons.
First, it has to cover effort.
Recital~26 of the GDPR names ``the costs of and the amount of time required for identification'' among the factors, and a trained model packages an inference that would otherwise take skill and compute, so a recipient might obtain cheaply what was costly inside.
This is covered if the entities inside are as free to use the model as the recipient is.
Second, it has to hold over time.
The guidelines note that a security incident may lead to a reassessment where the assessment relied on information being kept confidential (\S~42), and that the capabilities of entities may vary over time (\S~88).
The condition has to hold for as long as the recipient has the model, and released weights cannot be withdrawn if the recipient's situation changes, which is the concern of Question~\ref{q:5}.
Third, the recipient's information and means have to lie entirely within those available inside.
It is not enough that the recipient has less overall: a recipient may have less compute but different additional information, and what a third party knows cannot be listed in full.
Fourth, the premise has to cover everything the model was built from.
If several datasets are inside, the data has to be anonymous for the entities that hold them together.
The same applies to a pre-trained model, that is, an existing model that training starts from.
Its weights are held inside like any other data.
If they are anonymous for the entities inside, the argument covers a model built on them.
If they are not, for example because the pre-trained model has memorised personal data from its own training, the premise is not met and the argument does not apply.

The argument does not remove the need to document.
The guidelines ask controllers to document the anonymisation process, including the testing (\S~41).
In our reading, the assessment made for the data would be an input to the assessment of the model for its recipient, and would not replace it.

The guidelines address onward transfer of data: if it is reasonably likely that an entity can transfer the data to a third party, and ``it cannot be ruled out that this receiving entity has the means to identify the individual'', the data should be considered personal for that transfer (\S~21).
If the same holds for a model trained on the data, which is our reading, then what the release review has to establish is what \S~21 asks to be ruled out.
On this argument, the condition can be shown for a known recipient that works under comparable controls.
For a model that is published, the argument applies if the data is anonymous for any entity and not just for the entities inside, that is, if the data would itself pass under the simplified approach.
Where anonymity inside depends on what the entities inside do not know, as in Scenario~\ref{ex:3}, a published model reaches entities that know more, the telecom operator among them.
The argument then does not apply, and the model needs an assessment of its own, with the evidence described in Sect.~\ref{sec:n-framework}.
The argument thus shows under which condition a rule of thumb of the form ``a model trained on anonymous data is anonymous'' would hold; it does not confirm the rule in general.

\runin{Decentralised secure processing environments and federated learning.}
Sect.~\ref{sec:n-discussion} notes that the setting extends to decentralised environments, in which the data stays with each holder and partial results computed locally are combined~\cite{nou2025sardin}.
Federated learning is the case in which the partial results are updates to a shared model: each participant trains on its own data and sends an update to a server, and the combined model is sent back, over many rounds.
Such a federation is an environment in the sense of Sect.~\ref{sec:n-framework} to the extent that it controls who takes part, which processing is carried out and logged, and what each participant can bring to it and take out.

In the terms of Sect.~\ref{sec:n-framework}, an update that leaves a participant is an output, and over many rounds the updates amount to repeated releases of weights.
Attacks have been shown from the position of a server that sees the update of each participant, both to reconstruct training inputs~\cite{Zhu_DLG_2019} and to infer membership~\cite{zhu2025fedmia}.
Secure aggregation removes that position: the server learns the sum of the updates and not the individual ones~\cite{Bonawitz_SecAgg_2017}.
Br\"annvall et al.\ examine a variant in which the updates are combined under homomorphic encryption~\cite{SAIS23}, and noise can be added so that the sum carries a formal guarantee~\cite{McMahan_DPFL_2018,Kairouz_DPSecAgg_2021}.
The sum itself remains visible, to the server and, through the combined model, to the participants.
In our reasoning, how much it protects depends on how many contribute: with two participants, each can remove its own update and is left with the other's.

The Swedish data protection authority (IMY) examined federated learning between two healthcare providers in a regulatory sandbox~\cite{IMY_FedML_2023}.
It concluded that, under the conditions of that project, it could not be ruled out that the providers gain access to each other's personal data, since a party holding the jointly trained model and its own training data could attempt a membership inference or model inversion attack.
It stated that the conclusion does not hold for all forms of federated learning.

In the framework of this paper, federated learning calls for two assessments.
The first concerns the exchange among the participants, inside the decentralised environment: whether what one participant receives, the updates or the combined model, is anonymous for that participant, given its own data and means.
The IMY sandbox case is an assessment of this kind.
The second concerns the model that leaves the federation, which passes the release review and is assessed for its recipients as any other model (Release review, Sect.~\ref{sec:n-framework}).
For the assessment inside the federation, time enters as it does for a central environment: what has been exchanged stays with those who received it unless intermediate results are verifiably erased.
Where encrypted material is kept, the protection depends on whether the keys are destroyed and on whether the encryption will hold, which is among the foreseeable developments to consider (\S~29(e)).

\subsection{Sandboxes}\label{app:sandboxes}
Sect.~\ref{sec:n-discussion} concerns sandboxes in which personal data may be processed under a legal ground of its own, and asks whether a test protocol could be agreed with the authority in advance and whether a test of the anonymity of outputs falls within what Art.~59 of the AI Act allows.
This part adds detail on what a regulatory sandbox is, on Art.~59 and the European Health Data Space, on testing in a sandbox, and on Swedish experience.

\runin{What a regulatory sandbox is.}
The AI Act defines an AI regulatory sandbox as ``a controlled framework set up by a competent authority which offers providers or prospective providers of AI systems the possibility to develop, train, validate and test, where appropriate in real-world conditions, an innovative AI system, pursuant to a sandbox plan for a limited time under regulatory supervision'' (Art.~3(55))~\cite{EU_AI_Act_2024}.
The sandbox plan is a document agreed between the provider and the authority that describes ``the objectives, conditions, timeframe, methodology and requirements'' of the activities (Art.~3(54)).
Member States are to have at least one such sandbox operational by 2 August 2027, a date moved from 2 August 2026 by an amendment of July 2026~\cite{eu2026_1744}.

Burden and Stenberg compare the EU regulations that introduce regulatory sandboxes and find four features in common: the sandboxes support innovation, apply a regulation to something it has not yet been applied to, involve public authorities with explicit mandates, and operate under constraints that include a controlled setting and a limit in time~\cite{Burden_Stenberg_2025}.
They stress that a regulatory sandbox is not a loophole: exemption from requirements is a minor aspect in the regulations they compare, and the authority's opportunity lies in giving guidance in advance instead of enforcing afterwards.
The agreed plans are the instruments for governing a sandbox, making roles and responsibilities explicit.
The authors see a possible role for sandboxes in establishing first practice, recommend openness so that the documentation becomes available to a broad audience, and ask how far others, such as notified bodies, can take the outcomes into consideration.
They count as outcomes the reports, technical documentation, regulatory guidance and test results, and not the innovation itself.

A legal ground for further use of personal data is given by some of these regimes and not by others.
Burden and Stenberg find it in the AI Act and the Interoperable Europe Act, and not in the Cyber Resilience Act or the Net-Zero Industry Act.
The European Health Data Space, in their words, ``does not introduce a regulatory sandbox but opens for secondary use of personal data under contextual constraints''.
These are the regimes that Sect.~\ref{sec:n-discussion} refers to as a narrower group.

\runin{Article 59 of the AI Act.}
Art.~59 allows personal data that was lawfully collected for other purposes to be processed in an AI regulatory sandbox ``solely for the purpose of developing, training and testing certain AI systems'', when a list of conditions is met~\cite{EU_AI_Act_2024}.
The system has to be developed to safeguard a substantial public interest in one of the listed areas, among them public health, the environment, energy, transport and public administration (Art.~59(1)(a)).
The data has to be necessary for meeting requirements on high-risk systems that ``cannot effectively be fulfilled by processing anonymised, synthetic or other non-personal data'' (Art.~59(1)(b)).

Several conditions describe an environment like the one in Sect.~\ref{sec:n-framework}.
The personal data is to be kept ``in a functionally separate, isolated and protected data processing environment under the control of the prospective provider'', with access for authorised persons (Art.~59(1)(d)).
Logs of the processing are kept for the duration of the participation (Art.~59(1)(h)), and the data is deleted when the participation ends or the retention period is over (Art.~59(1)(g)).
On what may leave, the Act says that ``any personal data created in the sandbox cannot be shared outside the sandbox'' (Art.~59(1)(e)).

Two conditions concern documentation.
A ``complete and detailed description of the process and rationale behind the training, testing and validation of the AI system is kept together with the testing results'' as part of the technical documentation (Art.~59(1)(i)), and a short summary of the project is published by the authority (Art.~59(1)(j)).
At the end of a sandbox the authority provides an exit report on the activities and their results, which the provider may use ``to demonstrate their compliance with this Regulation through the conformity assessment process'' (Art.~57(7)).
The exit report can be made public if both parties agree (Art.~57(8)).
Where personal data is processed, the data protection authorities are to be associated with the operation of the sandbox (Art.~57(10)).

Two of these provisions bear on the questions of this paper.
We read point (e) as implying that what leaves the sandbox has to be shown not to be personal data, which is the task of a release review.
And point (i), with the exit report, gives a place where the evidence for the anonymity of outputs could be recorded.
Whether a test of that anonymity is covered by ``developing, training and testing'' is the legal question that Sect.~\ref{sec:n-discussion} leaves open.

\runin{The European Health Data Space.}
The Regulation on the European Health Data Space does not set up sandboxes, but it requires a secure processing environment for the secondary use of health data~\cite{EHDS_2025}.
Health data access bodies ``shall provide access to electronic health data pursuant to a data permit only through a secure processing environment'', which is subject to listed security measures, among them access for authorised persons named in the permit and logs of access and activities (Art.~73(1)).
On outputs, the access bodies ``shall review the electronic health data included in a download request to ensure that health data users are only able to download non-personal electronic health data, including electronic health data in an anonymised statistical format'' (Art.~73(2)).
This is a release review in the sense of Sect.~\ref{sec:n-framework}, set down in law.

\runin{Testing in a sandbox.}
Buscemi et al.\ observe that sandbox engagements ``vary considerably'', and report a distinction between those centred on legal and procedural oversight and those that also involve ``structured technical testing producing evidence as requested by'' the competent authority~\cite{buscemi2026configurator}.
From the conditions of such testing they derive eleven requirements for the infrastructure, and present an open-source framework that by their own assessment meets some of them in part.
One requirement is a tamper-evident log of data, code versions, parameters and results, since the authority ``requires evidence it can rely upon for the Exit Report and any subsequent regulatory follow-up''.

On who chooses the tests, the authors see ``an evident conflict of interest'' where the provider drives the selection: ``the party whose system is under scrutiny partly determines which evidence is produced, and an assessment can be made to look favourable through omission rather than misreporting''.
In the engagements they describe, the authority ``retains the authority to mandate a minimum set of tests''.
A consideration on the other side is that an authority which decides the tests takes on responsibility for that choice.
Sect.~\ref{sec:n-discussion} reports the arrangement and leaves open which is preferable.

The framework has a catalogue of tests with a planned tier of verified tests.
Before a test enters that tier, the authors would have its security and ``adequate documentation of what the plugin actually measures'' checked, and they leave open whether its claimed relation to legal requirements should be endorsed by an authority, and by which one.
The ten tests listed concern the safety and robustness of language models and the performance of other machine learning models; none concerns privacy.
The checks they describe do not include how reliably a test measures, which is the validation that our Question~\ref{q:1} asks about.

The AI Act has a requirement of its own on when the terms of a test are set.
High-risk systems are to be tested before they are placed on the market, and ``testing shall be carried out against prior defined metrics and probabilistic thresholds that are appropriate to the intended purpose of the high-risk AI system'' (Art.~9(8))~\cite{EU_AI_Act_2024}.
The requirement belongs to the risk management system, whose compliance is checked in conformity assessment (Art.~43), and it concerns the performance of high-risk systems and not anonymity.
Its principle, that metrics and thresholds are fixed before the test, is the one behind the question in Sect.~\ref{sec:n-discussion}: whether a test protocol for anonymity, with its attacks, metrics and thresholds, could be made part of the sandbox plan, which by its definition covers the ``methodology'' of the activities (Art.~3(54)).

An implementing regulation adopted in November 2026 lays down what a sandbox plan and an exit report contain at the least~\cite{eu2026_2487}.
The plan specifies, among other things, the objectives and scope of the activities, the legal requirements that are supervised, and ``an indication whether the AI system specified in the selected application involves the processing of personal data and the need for involving the competent data protection authority'' (Art.~5(2)).
Apart from a plan for testing in real-world conditions, where applicable (Art.~5(2)(g)), a test methodology, metrics and thresholds are not among the listed elements.
Since the list is a minimum, an agreed test protocol could still be made part of the plan.
The exit report serves regulatory learning and describes how the issues examined were addressed, ``including specific recommendations, good practices, tools, benchmarks or measures tested that could be replicated'' (Art.~6(3)).
It could record which test was used and what the test established, although it does not have the effect of a declaration of conformity (Art.~6(5)).

The AI Act also expects sandboxes to ``facilitate the development of tools and infrastructure for testing, benchmarking, assessing and explaining dimensions of AI systems relevant for regulatory learning'' (Art.~58(2)(i)).
Under the implementing regulation, authorities may share ``technical specifications, benchmarks and testing tools developed in AI regulatory sandboxes'' for publication, provided that the Commission has confirmed their ``proven relevance for compliance'' after consulting the AI Board (Art.~4(3))~\cite{eu2026_2487}.

A test of the anonymity of outputs, run on a model trained in the sandbox, is itself processing of the personal data inside.
Art.~59 allows processing for developing, training and testing an AI system, in the words quoted above.
On one reading, a test of whether the outputs of the system are anonymous is part of testing the system, and would then have a legal ground in the sandbox.
The main text leaves the question open, and so do we here.

\runin{Swedish experience.}
The Swedish data protection authority describes its sandbox as in-depth guidance to a specific innovation initiative on how the data protection rules should be interpreted and applied, given in workshops over some months and ending in a public report~\cite{IMY_FedML_2023}; the case of federated learning in A.2 was its pilot in 2022.
Within the collaboration programme eSam, three agencies and the data protection authority carried out a pilot of an AI regulatory sandbox as a discussion of one described system against questions drawn from the AI Act, without taking up the processing of personal data or testing~\cite{esam2024sandbox}.
A study of federated analysis of health data has been proposed for such a sandbox~\cite{SARDIN24}.
These are engagements centred on guidance, in the distinction reported above from Buscemi et al., and they did not include technical testing on personal data.

\subsection{Method validation}\label{app:validation}
Sect.~\ref{sec:n-discussion} treats a test result as a measurement with a detection limit, a bias and an uncertainty, asks that test methods are validated in advance on reference cases, and keeps the assessment of a release as a separate step.
Research on privacy attacks usually compares attacks with each other on benchmark data, which shows which attack is stronger.
Validation in the sense used here asks something else: what a given test could have detected on a given kind of release, and with what uncertainty.
This part adds detail on what is measured, on how a test can be weak, on validation and how far it carries, on validation in other regulated fields, and on indirect evidence.

\runin{What is measured.}
Leakage, as Sect.~\ref{sec:n-framework} defines it, is not observed directly.
A test reports a quantity that stands for it: the share of training records an attack identifies at a fixed low false positive rate (Sect.~\ref{sec:n-discussion}), the advantage of an attack over guessing~\cite{yeom2018overfitting}, or, in an audit, a lower limit on the privacy parameter (Appendix~\ref{app:dp}).
Metrology calls the ``quantity intended to be measured'' the measurand~\cite[2.3]{jcgm200}, and a test report should say which one it is.
Each of these quantities is the success of a particular attack, so, estimation error aside, none exceeds what the best attack would achieve.
The detection limit in Appendix~\ref{sec:app-detection} is stated for the first of them.
Bias, as Sect.~\ref{sec:n-discussion} uses the word, is a systematic difference between what the test reports and what the same attack would achieve on the release, for example because the tested records are not representative.
It is not a difference from a true leakage, which no test observes.

\runin{How a test can be weak.}
Sect.~\ref{sec:n-framework} names three reasons why a test may find no leakage: there is little leakage, the attack is too weak, or the test cannot detect the leakage for other reasons.
The literature gives examples of the second and third.

An attack can be too weak for the case.
For face and gait data, Hanisch et al.\ find that evaluations which assume an attacker unaware of the anonymisation ``grossly overestimate the performance of the anonymization''~\cite{hanisch2024false}.
For attacks that use reference models, the reference data matters: against a fine-tuned language model, Mattern et al.\ report that an attack identified 18.9\% of training records at a 1\% false positive rate when its reference model was trained on data from the same source, and 4.9\% when the data was similar but from another source~\cite{mattern2023neighbourhood}.
A tester who holds the original data avoids this (A.2).
The number of reference models matters as well.
In our own study, for one image benchmark, the same attack identified 12.9\% of training records at a 1\% false positive rate with 64 reference models.
With 8, the rate was 9.2\% when one variance was estimated for all records and 2.7\% when a variance was estimated for each record~\cite{brannvall2026bavaria}.
Reading such differences as a bias of the test, and not only as a weaker attack, is our interpretation.

A test can also report in a way that hides the records at risk.
Aerni et al.\ show for one case that the most exposed record was identified 99.9\% of the time at a 0.1\% false positive rate, where the figure for the whole dataset was 4\%~\cite{aerni2024misleading}.
Wang et al.\ find that different attacks, and repeated runs of the same attack, flag largely different sets of records, so that a single run ``may fail to capture the full spectrum of privacy leakage risks''~\cite{wang2025miatools}.
J\"alk\"o et al.\ show that applying one threshold to the pooled scores of all records hides false positive rates that differ widely from record to record~\cite{jalko2026reliability}.

The estimate itself can be off.
J\"alk\"o et al.\ also show that drawing the reference datasets from one fixed pool inflates the estimate, a bias in the other direction, and give a correction.
And low false positive rates take many runs to estimate: on the order of the inverse of the rate, according to Aerni et al., which is why the uncertainty is largest where the result matters most.

Stating what a test assumes is a first step against these weaknesses.
Salem et al.\ write the common attacks as games that make explicit what the attacker is given, how the data and the target record are chosen, and how success is measured~\cite{salem2023games}.

\runin{Validation on reference cases.}
Sect.~\ref{sec:n-discussion} asks that a test method is validated in advance, ``for its intended use, a class of releases and an attacker'', in reference cases where the leakage is known.
The phrase follows the meaning of validation in metrology: showing that the requirements a method meets ``are adequate for an intended use''~\cite[2.45]{jcgm200}.
A re-identification test is therefore not valid or invalid in general.
Its validation holds for a form of release, an assumed attacker and the decision it is to support, and a method validated for one use is not thereby validated for another.

Anonymeter gives an example of what a validation looks like.
Its authors evaluate the tool ``without relying on synthetic data since no synthetic data (or anonymization mechanism) can be assumed free of privacy risks''.
They build a release in which a known fraction of the records are copies of training records, and check that the reported risk rises with that fraction and is close to zero when nothing is copied~\cite{giomi2023anonymeter}.
The check also shows where the method does not behave as intended: for one of the three attacks, on two of the datasets, the reported risk did not follow the fraction of copies.
The tool reports its risk estimates with confidence intervals.

Whether a validation carries over to a given release is the difficult point.
A laboratory can validate a method on reference samples because the content of a reference sample is known independently of the method, and because the next sample behaves like the reference samples.
For an attack-based test neither is certain.
A reference case with known leakage has to be constructed, from copied records as in Anonymeter or from records inserted into training, so there is no independent reference value of leakage.
And how well an attack works depends on the model, the data and the training: the GRAIMATTER experiments found ``many hyperparameter values that appear safe for one dataset, but highly unsafe for others''~\cite{jefferson2022graimatter}, and Carlini et al.\ conclude that membership inference attacks ``only audit specific (dataset, model) tuples''~\cite{carlini2022privacyonion}.
Attacks also improve, so a method validated today is validated against the attacks of today.
The checks at each release that Sect.~\ref{sec:n-discussion} mentions address this in part: they ask whether the conditions under which the method was validated hold for the release at hand.

One answer is to place reference records in the release itself.
In an audit the tester knows which records were used in training, so membership is known; what is not known is how much the model leaks.
Records can be constructed to be among those the model is most likely to expose, and inserted into the training run.
Inserted records are known as canaries, after work that inserted randomly chosen sequences into the training data of language models to measure memorisation~\cite{carlini2019secret}.
If the test does not find these records, its failure to find others says little.
The share of them that it does find shows how sensitive the test is to such records in the model that is to be released (Appendix~\ref{sec:app-detection}).
Steinke et al.\ show that such an audit can be carried out with a single training run, at the price of lower limits that are less tight than those of audits with many runs~\cite{steinke2023one}, and Aerni et al.\ design the inserted records ``to mimic the most vulnerable samples in the data''~\cite{aerni2024misleading}.
What remains an assumption is that the real records at risk behave like the inserted ones.

\runin{Validation in other regulated fields.}
Question~\ref{q:1} names fields in which a method has to be validated before its results support a decision.
The requirements are given here in more detail.

In metrology, a result ``is complete only when accompanied by a statement of the uncertainty''~\cite[3.1.2]{jcgm100}, and validation is tied to an intended use, as noted above.
ISO/IEC 17025 asks a testing laboratory to evaluate the uncertainty of its measurements and to validate methods that are not standard methods~\cite{iso17025}.
When a result is compared with a limit, a decision rule states ``how measurement uncertainty will be accounted for with regard to accepting or rejecting an item''~\cite[3.3.12]{jcgm106}.
The same guide defines the risk of accepting an item that does not conform and the risk of rejecting one that does, and shows how the first can be reduced by setting the acceptance limit inside the permitted range.

In banking, institutions that use internal ratings are to ``have robust systems in place to validate the accuracy and consistency'' of their rating systems, and to compare realised default rates with the estimated probabilities (Art.~185)~\cite{crr2013}.
US supervisory guidance describes as sound practice an ``effective challenge'' by individuals with ``sufficient independence to maintain objectivity'', and outcomes analysis that compares model outputs with real-world outcomes, of which back-testing is one form~\cite{sr262}.
The Basel framework for market risk models is a close analogue to a re-identification test.
It counts the exceptions in a sample of 250 observations and reports the probability of rejecting an accurate model and of accepting an inaccurate one.
Since ``there is no threshold number of exceptions that yields both a low probability of erroneously rejecting an accurate model and a low probability of erroneously accepting all of the relevant inaccurate models'', it uses three zones, and in the middle one the supervisor ``should encourage a bank to present additional information about its model before taking action''~\cite{bcbs_mar99}.

Other fields state the requirements in numbers.
In vitro diagnostic devices are to achieve the analytical performance stated by the manufacturer, such as trueness, precision and ``limits of detection and quantitation'' (Annex~I, 9.1(a))~\cite{ivdr2017}.
For residues in food-producing animals, a screening method may be used if it is validated and has ``a false compliant rate lower than or equal to 5~\%'', and the rules define a detection capability as ``the smallest content of the analyte that may be detected or quantified in a sample with an error probability of~$\beta$''~\cite{eu2021_808}.
For drinking water, methods are validated in accordance with ISO/IEC 17025, the limit of quantification is to be ``30~\% or less of the relevant parametric value'', and the uncertainty of measurement ``shall not be used as an additional tolerance''~\cite{dwd2020}.
The monitoring of water status sets comparable criteria~\cite{eu2009_90}.

These requirements have three things in common: the method is validated apart from its use, what it can detect is stated with the probability of error, and the rule for deciding under uncertainty is set before the result is known.
A re-identification test used to show that leakage stays below a threshold meets the same questions, and Appendix~\ref{sec:app-detection} works through the first two for a simple case.

\runin{Traceability and decision rules.}
Traceability and decision rules are not terms of the guidelines, but they describe what the test report in Sect.~\ref{sec:n-discussion} asks for.
In metrology a result is traceable when it ``can be related to a reference through a documented unbroken chain of calibrations, each contributing to the measurement uncertainty''~\cite[2.41]{jcgm200}.
For a re-identification test the chain would run from the measurement on the release, through the validated method, to reference cases with known leakage, with the data tested, the attacker assumed and the protocol documented at each step.

A decision rule says how the uncertainty of a result is used when the result is compared with a limit.
One form is guarded acceptance, in which the acceptance limit is set inside the permitted range so that ``the risk of accepting a non-conforming item can be reduced''~\cite[8.3.2.1]{jcgm106}.
For a release, this would mean comparing the upper confidence bound of the measured leakage with the threshold, and not the estimate itself.
The rule protects against the uncertainty of the measurement, at the cost of rejecting some releases that are anonymous.
It concerns the tested attack only, since a stronger attack may find more.
The three zones of the Basel framework are another form, in which a result in the middle zone calls for more information.
Which threshold applies cannot be settled by measurement (Sect.~\ref{sec:n-conclusion}); the decision rule concerns how a measurement is set against it.

Reporting of uncertainty is still developing in tools for privacy testing.
The GRAIMATTER recommendations ask for confidence intervals where possible (A.2), and the open-source tool LeakPro reports uncertainty estimates for some of its metrics~\cite{leakpro_repo}.
Among other open-source tools that run membership inference attacks, ML Privacy Meter, an early example, reports risk scores for individual records and for the dataset as a whole; the 2020 paper that describes it does not report the uncertainty of its results or a validation of them~\cite{murakonda2020mlprivacymeter}.

\runin{NIST guidance on testing.}
Of the publications named in Sect.~\ref{sec:n-discussion}, two from NIST discuss testing.
NIST SP 800-188 asks agencies to document that ``the tests that were used to evaluate the effectiveness of the de-identification were validated for that purpose'', and lists motivated intruder tests among the ways to validate privacy protection~\cite[Sects.~3.3 and~4.6]{nist800188}.
NIST SP 800-226 notes that audits cannot prove that a guarantee holds (Appendix~\ref{app:dp})~\cite[Sect.~2.7]{nist800226}.
Neither says how the validation of a test is carried out.

\runin{Who validates.}
The fields compared above have arrangements around validation that go beyond the single laboratory.
References are maintained to which results can be traced, and laboratories accredited under ISO/IEC 17025 monitor their performance by comparing results with other laboratories~\cite{ilacp9}.
We are not aware of comparable arrangements for re-identification tests.
Three things would be needed: reference cases with known leakage that are openly available, so that methods can be validated and compared on the same cases; comparisons between tools on such cases; and a body with a mandate to vet tests, a question that Buscemi et al.\ also leave open (A.3).
There are beginnings: NIST states that it ``is actively working on empirical privacy metrology''~\cite[Sect.~2.2]{nist800226} and runs testbeds in which privacy risks are to be evaluated ``using a set of empirical privacy evaluation metrics''~\cite{nist_pets_testbeds}, and the AI Act expects regulatory sandboxes to help develop tools for testing (A.3).
Appendix~\ref{app:staged} names parties that have reason to validate.

\runin{Indirect evidence and risk indicators.}
Question~\ref{q:7} asks for which releases indirect evidence can be sufficient.
In Scenario~\ref{ex:7} a hospital argues from three indicators: a small model, strong regularisation, and almost the same performance on training and test data.

Guidance already reasons from indicators.
For data, the guidelines list key factors, such as whether the data is aggregated, how much information there is on each individual and how many individuals are included, and state ``as a rule of thumb'' which data re-identification techniques are more likely to succeed against (\S\S~85--86).
For models, Opinion 28/2024 lets supervisory authorities consider whether the training ``uses regularisation methods to improve model generalisation and reduce overfitting'', and documentation of ``the ratio between the amount of training data, and the number of parameters in the model''.
It adds that the presence or absence of such elements ``is not a conclusive criterion''~\cite[\S\S~48, 52, 58]{edpb2024opinion28}.

There is a reason why the gap between training and test performance says something about leakage.
Yeom et al.\ show that for a classification model a simple attack gains an advantage equal to that gap, so that ``high generalization error necessarily results in privacy loss''~\cite{yeom2018overfitting}.
The converse does not hold: in their words, ``overfitting is not necessary'' for an attack to succeed.
A large gap is therefore evidence of leakage, while a small gap indicates a lower risk without ruling leakage out.
The gap is also an average, and a small average does not rule out a few exposed records (Question~\ref{q:2}).

Work on output checking reaches the same conclusion for settings in general.
GRAIMATTER names kinds of model that carry the training records inside them and are ``generally not appropriate for release'', and suggests simple checks, for example that a model file ``ought to be orders of magnitude smaller than the size of the training data''.
On how far such knowledge goes it says: ``While we can know that some hyper-parameter/algorithm variants typically lead to disclosive models, it is in the nature of logic that we cannot know the converse in advance for most algorithm-dataset combinations''~\cite{jefferson2022graimatter}.
The SACRO-ML checks before training apply this in one direction, warning about settings outside limits set by the environment (A.2).

Two further examples show indirect evidence in use.
Findata asks holders of a data permit who want to export a model to submit a form of 28 questions on the model, privacy-enhancing techniques, intended use, training data, performance and anonymity.
One question asks whether resistance to attacks has been tested; the authority then ``verifies the anonymity of the results through a risk-based assessment''~\cite{findata2025anonymous}.
Jebreel et al.\ define five conditions that must all hold for a membership inference attack to be a real threat, among them that it works on a model that is not overfitted and that it identifies members at a false positive rate near zero, and find that none of the attacks they review meets them all~\cite{jebreel2026critical}.
Sect.~\ref{sec:n-points} suggests that such conditions could serve as a check on a release.
That use is ours; the authors propose the conditions for judging attacks.

Taken together, indicators can show that a release is probably unsafe, and can support a judgement that it is probably safe, but the sources above do not treat them as establishing safety.
Sect.~\ref{sec:n-discussion} suggests one way forward: validated tests on reference cases could show how well an indicator predicts what a test finds.
An indicator used in that way resembles a screening method in the fields compared earlier, where a result that raises suspicion is followed by a confirmatory method, and it is the first stage of the staged assessment in Appendix~\ref{app:staged}.

\section{The detection limit: a worked example}\label{sec:app-detection}
This appendix works through the example behind Scenario~\ref{ex:1}: a test that wrongly flags one record in a thousand, applied to 1,000 training records.
It shows how many flags are needed before a result differs from chance, how much the test can reliably detect, and what a result says once the test has been run.

\runin{The test.}
A membership inference attack scores each record and flags it if the score exceeds a threshold fixed in advance.
We apply the attack to $N$ training records and count the flagged records, the hits $X$.
The false positive rate $f$ is the probability with which the attack flags a record that was not used in training.
We start from $N=1000$ and $f=0.001$, with $f$ known, and relax these choices below.

If the model leaks nothing, a training record is flagged with the same probability $f$ as a record that was not used.
Assuming that records are flagged independently of each other, $X$ then follows a binomial distribution with mean $Nf=1$.
Then $P(X\geq 3)=8.0\%$ and $P(X\geq 4)=1.9\%$.
We call a result different from chance if a result at least as large would occur in fewer than 5\% of tests of a model that leaks nothing, a level commonly used in statistics; by this rule the result differs from chance only from four hits upwards.
Because counts are discrete, the actual error probability of this rule is 1.9\%.
This decision threshold tells us how to read a result, but not what the test is able to find.

\runin{Detection limit.}
Let $q$ be the probability that a training record is flagged, so that $X$ is binomial with parameters $N$ and $q$.
This is the quantity the test estimates: the share of training records the attack identifies at the false positive rate $f$ (A.4).
The detection limit is the smallest $q$ for which $P(X\geq 4)\geq 95\%$.
This follows the definition used in metrology, where the detection limit is the value at which the probability of wrongly claiming absence is small, given the probability of wrongly claiming presence~\cite[4.18]{jcgm200}.
In statistical terms it is the result of a power calculation for a one-sided test.
Here the limit is $q=0.77\%$, or 7.7 expected hits, of which about one arises by chance.

To read the limit as a statement about leakage, a model of leakage is needed.
The simplest is that a share $\pi$ of the records leak and the others do not, and that the attack flags a leaking record with probability $r$, its sensitivity.
Then $q=\pi r+(1-\pi)f$.
Exposure is in reality a matter of degree, and the division into records that leak and records that do not is made to keep the example readable.
With $r=1$ the limit corresponds to $\pi=0.67\%$, about seven records.

If instead exactly $k$ records leak and $r=1$, then $X=k+Y$, where $Y$ counts the false positives among the other records. The probability of reaching four hits is 8\%, 26\%, 63\% and 100\% for $k=1,\dots,4$, so four leaking records suffice. The two readings differ because the number of leaking records is itself random in the first. The difference shrinks once $r<1$: for $r=0.5$ a fixed number requires 12 records and a share requires 1.35\% (13.5 records), and for $r=0.1$ the figures are 67 records and 6.8\%.

The limit grows roughly as $1/r$: 0.67\% for $r=1$, 1.35\% for $r=0.5$ and 6.8\% for $r=0.1$. The sensitivity is unknown until the method has been validated on cases with known leakage, for example with canaries, so the limit cannot be stated as leakage without such validation (A.4, Validation on reference cases).
In plain terms, a handful of leaking records can be detected if the attack finds every one of them, and many more are needed if it finds only some.

\runin{Sample size and false positive rate.}
Table~\ref{tab:detection} varies $N$ and $f$ for $r=1$. With no false positives, a single hit suffices and the limit is about $3/N$. A false positive rate well below $1/N$ gains little. A larger test detects a smaller share of the records, but the corresponding number of records grows.

\begin{table}[t]
\centering
\caption{Decision threshold and detection limit, given as the share of leaking records for $r=1$, with $f$ known, 5\% significance and 95\% detection probability.}
\label{tab:detection}
\begin{tabular}{rlccc}
\hline
$N$ & $f$ & Hits needed & Actual error & Detection limit \\
\hline
1,000 & 0.01 & 16 & 4.8\% & 1.31\% (13.1 records) \\
1,000 & 0.001 & 4 & 1.9\% & 0.67\% (6.7 records) \\
1,000 & 0.0001 & 2 & 0.5\% & 0.46\% (4.6 records) \\
1,000 & 0 & 1 & 0 & 0.30\% (3.0 records) \\
10,000 & 0.001 & 16 & 4.9\% & 0.13\% (13.1 records) \\
100,000 & 0.001 & 118 & 4.3\% & 0.036\% (36.4 records) \\
\hline
\end{tabular}
\end{table}

In practice $f$ is estimated. If one of 1,000 non-members is flagged and the members are compared with the non-members in a one-sided Fisher exact test, seven member hits are needed at the 5\% level, and over repeated experiments with $f=0.001$ and $r=1$ the detection limit rises to 1.2\%.

\runin{After the test.}
The detection limit describes the test before it is run. A result gives an upper confidence bound instead. With $f$ known and one hit observed, the 95\% upper bound on the mean number of hits is 4.7. Subtracting the one expected by chance leaves 3.7 hits, or 0.37\% of the records for $r=1$ and more for lower sensitivity. With no hits and $f=0$ the bound is 0.30\%. Either bound concerns the tested attack only, since a stronger attack may have a higher sensitivity.
Comparing such an upper bound with a threshold is the decision rule described in A.4.

The decision threshold and the detection limit correspond to the decision limit CC$\alpha$ and the detection capability CC$\beta$ in the regulation on residues in food-producing animals~\cite{eu2021_808} (A.4). The calculation assumes independent records, the same false positive rate for every record, and a single attack with one threshold; choosing among several attacks or thresholds after the fact raises the actual error probability. Expressed in individuals, a test of this size can leave several people in a thousand exposed without showing it reliably, while the guidelines require an insignificant likelihood of re-identification for all individuals (\S\S~36, 87).

\section{A staged assessment and the test report}\label{app:staged}
Sect.~\ref{sec:n-discussion} (Method validation) mentions that a release can be assessed in stages and that the result should be documented in a test report.
This appendix describes one way of doing so for a trained model; the attacks and measures it names are examples.
We call the party that carries out the assessment the assessor, a word the guidelines also use (\S~47).

\runin{A staged approach.}
An attack-based test of a trained model can be expensive.
Many established attacks rely on reference models, which the assessor trains in the same way as the model under test in order to learn how such a model behaves on records it was and was not trained on, and an attack may need many of them.
An assessment can therefore begin with a pre-screening that requires little computation and shows how much testing the release needs.
An attack test follows where it is needed, and the results are documented in a test report.
If the test shows too much risk, or if circumstances later change, the release is assessed again.
All of this presupposes a test method that has been validated beforehand, which is done separately so that the work is not repeated for every release.
Table~\ref{tab:protocol} gives an overview.

\begin{table}[t]
\centering
\caption{Example of a staged assessment.}\label{tab:protocol}
\scriptsize
\begin{tabular}{@{}L{2.4cm}L{5.3cm}L{3.8cm}@{}}
\hline
\textbf{Part} & \textbf{What is done} & \textbf{When}\\
\hline
Method validation & The detection limit, bias and uncertainty of the test method are established on reference cases & Beforehand, apart from any single release\\
Pre-screening & The assessor records what is assessed and against whom, fixes the protocol, and examines documentation, risk indicators and any formal guarantee & For every release\\
Attack test & The assessor runs the attacks of the protocol, checks that the conditions of the validation hold, and compares with other attacks & Where the pre-screening shows that more evidence is needed\\
Test report & The assessor documents the method, the results and their limits for those who decide on the release & For every release\\
Re-assessment & The release is assessed again, possibly with additional mitigations applied, and each round is reported & When a test shows too much risk, or when a condition for reassessment occurs\\
\hline
\end{tabular}
\end{table}

\runin{Validation of the method.}
Validating a test method means establishing its detection limit, bias and uncertainty on cases where the leakage is known: the smallest leakage it reliably finds, whether it over- or underestimates, and how certain its results are (A.4, Validation on reference cases).
A method validated for a kind of release and for stated assumptions about the attacker should then be usable in every assessment that fits them.
Several parties have reason to validate, although no settled arrangements exist (A.4, Who validates).
The maintainers of a testing library may want to show under which assumptions their tool is expected to work, and an organisation that offers assessments may want to document how its methods were validated, so that clients and authorities can judge what its results are worth as evidence.
An organisation can also set up a validation function of its own, preferably one that is independent of those who develop the model or own the product, as model validation is in banking (A.4).
A validation could also be carried out under the supervision of an authority in a regulatory sandbox (A.3).
Because a validation holds only for the conditions under which it was made, these conditions need to be documented with it.
When a method is applied to a new kind of model or task, they may no longer be met, and the affected parts of the validation then have to be repeated.

\runin{Pre-screening.}
The first stage is a pre-screening of the risk, limited to what can be done with little effort; in practice this means without attacks that require substantial computation.
The assessor first records what is assessed and against whom: the form of the release, who will receive it, and what information and means an attacker is assumed to have (Sect.~\ref{sec:n-framework}).
In Scenario~\ref{ex:4}, for example, the company chooses between shipping the weights of its model and offering the model through a query service.
For the weights, a test would have to assume an attacker who holds the model and can query it without limit; for the service, it could assume an attacker bound by the limits of the service, with a margin because such limits can be circumvented (Sect.~\ref{sec:n-discussion}, Model weights).

The assessor then examines the indirect evidence.
The developer's documentation describes the type and size of the model, the amount of training data, who will have access, and the protective measures applied in training, such as regularisation, which keeps a model from fitting its training data too closely, or differential privacy.
The questionnaire of Findata, the Finnish authority for the secondary use of health data, asks for information of this kind before a model may leave its environment (Question~\ref{q:7}).
The assessor can also compute risk indicators from the model, for example the generalisation gap: the difference between the model's performance on its training data and on data it has not seen.
A large gap is evidence that the model has retained details of individual records; a small gap indicates a lower risk but does not rule leakage out (A.4, Indirect evidence and risk indicators).
Training with differential privacy, if implemented correctly, gives a guarantee: an upper limit on leakage that holds for any attack.
With small privacy parameters the guarantee may itself support the conclusion that leakage is low (Question~\ref{q:6}).
Large models are commonly trained with large parameters, and the limit then says less, so an attack test may be warranted.

The assessor then decides whether an attack test is needed and how extensive it should be, in proportion to the risk and the intended use of the release (Sect.~\ref{sec:n-discussion}, Method validation).
In Scenario~\ref{ex:7} the hospital argues that the pre-screening is enough, because its model is simple and strongly regularised and the generalisation gap is small.
Whether indirect evidence can be sufficient in this way is Question~\ref{q:7}.
If a test is to be run, its protocol is fixed before it starts: the attacks, the measures, and the criteria against which the results will be judged.
Choosing among attacks or thresholds after the results are known would make a favourable result more likely to arise by chance (Appendix~\ref{sec:app-detection}).

\runin{The attack test.}
We use membership inference as the example of an attack test, as in the rest of the paper.
The attack predicts for each record whether it was used in training, and the assessor compares the predictions with the truth, which requires records that were used in training and similar records that were not.
Such a test is easiest to arrange before release, as in Scenario~\ref{ex:3}.
Inside the environment where the model was trained, the assessor has the training records and can set similar records aside before training (A.2, Testing inside the environment).
The result is commonly given as the true positive rate at a low false positive rate, that is, the share of training records the attack identifies while wrongly flagging only a very small share of the others.
It is reported with its uncertainty, and also for the most exposed records, which an average can hide (A.4, How a test can be weak).
Other kinds of attack have other measures, and for some models another attack may matter more, such as the extraction of training text from a language model~\cite{carlini2021extracting}.

Where reference models can be afforded, the main attack would be a validated attack from the current state of the art, such as LiRA~\cite{carlini2022first} (A.1), so that the result can be reproduced and compared.
Where they cannot, for instance for a very large model, a simple attack may be the only feasible test.
For classifiers, the attack of Yeom et al.\ is an example: it predicts that a record was used in training when the model's error on it is small~\cite{yeom2018overfitting}.
If such an attack identifies many training records, this indicates leakage.
A low result allows no conclusion, since a stronger attack may find more, and the assessment then rests more on the other types of evidence.
Duan et al.\ describe this situation for large language models.
Attacks that need many reference models are often intractable there, and the five simpler attacks they evaluated did little better than guessing in most settings, on the data used to pre-train such models; the authors add that stronger attacks might do better~\cite{duan2024mia}.

The assessor also checks that the conditions of the validation hold for the model at hand.
For LiRA, three examples can be given.
The reference models should resemble the model under test, for instance in accuracy, in generalisation gap and in how well the attack works against them.
The normal distribution that LiRA assumes for the outputs of the reference models should fit; our study of established attacks found that it does for some datasets and models but not for others~\cite{brannvall2026bavaria}.
And the number of reference models should be within the range for which the method was validated: our study also found that results change with this number, and that simpler attacks are more dependable than LiRA when only a few are available (A.4, How a test can be weak).
These checks serve as quality criteria.
The assessor reports the deviations found and, where a deviation is large, considers whether the detection limit and uncertainty from the validation still apply.

Running further attacks on the same model helps to interpret the result, since a low number can mean little leakage or a weak attack.
A simple attack shows what is achieved with little effort.
At the other extreme, an attack can be given more information than a realistic attacker would have.
The SACRO-ML tools include a ``worst-case'' attack that is trained on the model's outputs for records whose membership is known~\cite{smith2025sacroml}, and in ongoing work we have experimented with a variant of the RAPID attack~\cite{he2024rapid} that is trained in the same way.
If the main attack identifies no more than the simple one, the assessor first goes back to the checks above.
A large distance to the worst-case attack may indicate that a better-informed attacker could achieve more, although an attack trained with known membership can overstate what any real attacker achieves.

For a model trained with differential privacy, the test can take the form of an audit, which gives a lower limit on the privacy parameter.
A lower limit above the guaranteed value points to an error in the implementation or the calculation, and one below it indicates how much room the guarantee may leave, as far as the audit uses a strong attack (Appendix~\ref{app:dp}).

\runin{The test report.}
The assessor writes the report for those who decide on the release and for anyone who later reviews the decision, such as a supervisory authority.
Such a reader may be unable to repeat the test, in particular once the data has been deleted, so the report needs to show what the result is worth as evidence.
The guidelines place the documentation of the anonymisation process, including the testing, with the controller (\S~41).
A report could contain, for example, the following items.
\begin{itemize}
\item What was assessed: the form of release, the controls it relies on, and the attacker assumed, and whether the assumptions match the expected attacker or are stronger (Sect.~\ref{sec:n-framework}; A.1).
\item Who carried out the test and when, and whether the validation or the test was carried out or reviewed independently (Sect.~\ref{sec:n-discussion}).
\item The protocol fixed before the test: the attacks, measures and criteria, the records to be tested and how the comparison records are obtained (Sect.~\ref{sec:n-discussion}; A.2; A.3).
\item What the test could detect: the quantity measured, the detection limit, the lowest false positive rate that the available records can estimate, the validation of the method and the outcome of the checks (A.4; Appendix~\ref{sec:app-detection}).
The detection limit depends on the validated method and on the number of records in the test, and stating it lets the reader see when a finding of low risk rests on a test with limited ability to detect leakage.
\item The results with their uncertainty, including results for the comparison attacks and for the most exposed records (Sect.~\ref{sec:n-discussion}; A.4).
The latter matter in a case such as Scenario~\ref{ex:2}: when a data subject requests erasure after release and the training data is gone, a report with only an average says very little about that person.
\item Any formal guarantee with its parameters, unit of privacy and variant, the mitigations applied, and the results of earlier rounds (Appendix~\ref{app:dp}).
\item Results for stated prior beliefs.
Footnote~44 of the guidelines asks whether the given data yields a ``significantly different `posterior belief'\,''.
What an attacker believes after the attack has flagged a record depends on what the attacker believed before, so the report can give the belief after a flag for a few prior beliefs, as Appendix~\ref{app:dp} does for a differential privacy guarantee.
Jayaraman et al.\ propose to report this as the share of flagged records that really were training records, and note that the usual test design, in which half of the candidate records are training records, corresponds to a prior that is rarely realistic~\cite{jayaraman2021revisiting}.
In their experiments on models trained without differential privacy, this share fell for three of the four attacks as training records became a smaller part of the candidates, while the strongest attack still identified some records with high confidence.
Which change in belief is significant cannot be settled by measurement (Sect.~\ref{sec:n-conclusion}).
\item The conditions for reassessment (cf.\ \S~35 of the guidelines).
\end{itemize}
A report is also written when the assessment stops after the pre-screening; it then records the indirect evidence and why no attack test was considered necessary.

\runin{Re-assessment.}
A release is assessed again in two situations.
The first is when a test shows more leakage than is acceptable; what level is acceptable cannot be settled by measurement (Sect.~\ref{sec:n-conclusion}).
The developer can then retrain the model with added protective measures, such as stronger regularisation or differential privacy, or choose a more restricted form of release, and the new release is assessed again; running the same test before and after shows what a measure has achieved (Sect.~\ref{sec:n-discussion}).
Every round is reported and the mitigations added are documented, so that the final result is not read as if it were the only attempt.
Where anonymity cannot be shown, one option is to handle the release as personal data under the GDPR; what that requires is outside this paper.
The second is when a condition for reassessment named in the report occurs, such as the publication of a stronger attack or a change to the model or to the access given to it (\S~35).
The AI Act describes risk management for high-risk AI systems in similar terms, as ``a continuous iterative process planned and run throughout the entire lifecycle'' (Art.~9(2))~\cite{EU_AI_Act_2024}.
In Scenario~\ref{ex:5} nothing can be withdrawn when a stronger attack appears, but a report that states which attacks were tested, and under which assumptions, shows whether the new attack goes beyond what was covered.

\runin{In practice.}
An organisation that first asks whether a model is anonymous when the model is ready to leave may learn too late that it is not.
By then the model is built and its release expected, and what remains is to retrain it or to restrict the release.
The questions of the pre-screening can be asked while the model is being developed, on documentation and risk indicators alone, when the design can still be changed at little cost.

When a test is run, the organisation needs to know what the test could have found.
A result of ``no leakage detected'' on a thousand records is compatible with several individuals in a thousand being exposed (Appendix~\ref{sec:app-detection}), while the guidelines require an insignificant likelihood of re-identification for every individual (\S\S~36, 87).
A report that does not say what the test could detect leaves the decision maker with nothing to weigh, which is why Sect.~\ref{sec:n-conclusion} recommends validated methods and results reported with their uncertainty.

The test needs the training data and comparable records, so it has to be planned and carried out before the data is deleted (A.2).
The test is itself processing of the original data, which Sect.~\ref{sec:n-discussion} (Airlocks) takes up.
Once the data is deleted, the documentation of the test is what remains, and the guidelines ask that it is retained (\S~41).
The test should be run by a function with sufficient independence from those who develop the model and from the release decision, as supervisory guidance describes for model validation in banking~\cite{sr262}.

Much of this is existing practice.
Findata asks for documentation before a model may leave its environment, and the SACRO-ML tools check the settings of a model before training and run attacks on it afterwards (Question~\ref{q:7}).
The arrangement described here also validates the method separately, fixes the protocol in advance, and reports what the test was able to detect together with every round of assessment.

\section{More on differential privacy}\label{app:dp}
The Background briefly introduces differential privacy, Question~\ref{q:6} asks under what conditions such a guarantee can support an anonymity assessment, and Sect.~\ref{sec:n-discussion} sets a guarantee beside an attack-based test.
This appendix gives the detail behind those passages: what a stated guarantee means, how it relates to the comparison in the guidelines, and why a guarantee and a test complement each other.

\runin{Differential privacy in more detail.}
Differential privacy is a property of the process that produces a release.
A process here is whatever turns the original data into the output: computing a table of statistics, answering a query, or training a machine learning model.
In the notation of the guidelines it is the anonymisation process A, which produces the given data A(D) from the original data D (Fig.~\ref{fig:inference-test}).
Differential privacy requires that the chance of any outcome is about the same whether or not one individual's data is included~\cite{dwork2014algorithmic}, \cite[Sect.~2.1]{nist800226}.
This is achieved by building randomness into the process, for example noise added to a statistic or to the training steps of a model.
What a stated guarantee means depends on several choices, which NIST describes in its guidance for those who evaluate such guarantees~\cite{nist800226}.

The first is the privacy parameter $\varepsilon$, which sets the limit: a smaller $\varepsilon$ means stronger protection and lower accuracy.
NIST offers loose starting points.
It reports from one study that $\varepsilon$ of 0.1 generally provides strong protection and that values below 1 are considered reasonable, and notes that many deployments have used values between 1 and 20 and that values at the upper end of that range do not always provide meaningful protection in practice~\cite[Sect.~2.2]{nist800226}.
Many guarantees carry a second parameter, $\delta$, a small probability with which the limit may fail, possibly completely, which is why $\delta$ is set very small~\cite[Sect.~2.3]{nist800226}.

The second choice is the unit of privacy: what counts as ``the data of one individual''.
A guarantee per record protects each record, and a person who contributes many records is protected less than the stated $\varepsilon$ suggests~\cite[Sect.~2.4]{nist800226}.
The third is the variant of differential privacy.
Several are in use, each with its own parameters, and converting them to a common form loses precision, so NIST asks that the original parameters are reported~\cite[Sects.~2.3 and~2.5]{nist800226}.

Guarantees also add up.
When the same data is analysed several times, the parameters of the analyses can be added to give an upper limit on the total, often called the privacy budget~\cite[Sect.~2.2]{nist800226}.
This is the formal counterpart of the point in \S~77 that several outputs from the same data can reveal more than each does alone.
Open-source libraries for training with differential privacy, such as Opacus, keep account of the budget spent as training proceeds~\cite{yousefpour2021opacus}.

The guarantee has limits.
Differential privacy does not prevent inferences about an individual that would be drawn in the same way without that individual's data~\cite[Sect.~2.1]{nist800226}, which matches the inference from the population at large that the guidelines accept (\S~70).
Where the data of different individuals is correlated, as for members of a family, the data of others can reveal information about a person whom the guarantee was meant to protect~\cite{kifer2011nofreelunch}.

How such a guarantee relates to legal standards has been studied for particular cases.
Nissim et al.\ extract a conservative mathematical model of the privacy requirements of a US education privacy law and prove that differential privacy satisfies it, leaving open how the parameter should be set~\cite{nissim2018bridging}.
Cohen and Nissim formalise one aspect of the GDPR's notion of singling out and prove that differential privacy protects against it, whereas k-anonymity typically does not; they regard such protection as necessary but not sufficient for anonymisation~\cite{cohen2020singling}.
Cohen et al.\ derive a small set of technical properties that any anonymisation rule should satisfy, and find that prevailing interpretations of singling out, linkability and inference fail some of them~\cite{cohen2024regulations}.
On the size of the parameter, Domingo-Ferrer et al.\ caution that in data release and machine learning, acceptable accuracy is typically obtained only with values of $\varepsilon$ for which the guarantee is no longer meaningful~\cite{domingo2021limits}.

\runin{The comparison in the guidelines.}
The Framework says that formal and empirical evidence rest on the comparison that the guidelines describe.
The literature on differential privacy supports this.

Footnote~43 sets what can be inferred about a person from A(D) against what could be inferred from A(D$'$), produced without that person's data, and footnote~44 asks whether the given data yields a ``significantly different `posterior belief'\,''.
Differential privacy has a formulation of the same kind: for a small $\varepsilon$, whatever an observer knew beforehand, the belief formed after seeing the output is close to the belief the observer would have formed had the individual's data been left out~\cite{kasiviswanathan2014semantics}.
Wood et al.\ explain this reading for a non-technical audience and show how $\varepsilon$ limits the change in belief~\cite{wood2018primer}.

An attacker who tries to decide whether an individual's data was included is carrying out a statistical test, and with a small $\varepsilon$ the guarantee means that the test cannot have both a low false positive rate and a low false negative rate~\cite{wasserman2010statistical,kairouz2015composition}.
Later work states the guarantee directly as a trade-off between the two error rates~\cite{dong2022gaussian}, and NIST points to this reading for interpreting and comparing guarantees~\cite[Sect.~2.3]{nist800226}.
A membership inference attack is such a test carried out in practice.
In the terms of footnote~43 it is the case where the property S is whether the person's data was in D (Fig.~\ref{fig:inference-test}).

In our reading, the guarantee and the comparison in the guidelines still differ in three respects.
Take the John Smith example from footnote~43 and a model trained on data that includes his.
First, the comparison concerns one case: this model, John Smith, and a given inference about him.
The guarantee is a statement about the training process and holds for all datasets, persons and inferences.
It is broader than the comparison requires, although how strong it is depends on its parameters.
Second, if the model trained with John Smith's data can hardly be told apart from one trained without it, then what is inferred from the model cannot depend much on his data, so a limit on telling the two apart covers other inferences and attacks as well.
A test tries one attack on one property, for example whether John Smith was in the training data.
If it fails, a stronger attack, or an attack on another property, may still succeed.
Third, $\varepsilon$ limits how far a belief about John Smith can move because his data was used.
It does not say whether a movement is ``significant'' or one ``on which an action could turn'' (footnote~44).
With $\varepsilon = 1$, for example, a belief of 1\% can rise to at most 2.67\% and a belief of 50\% to about 73\%~\cite{wood2018primer}.
Whether either change is significant depends on what is at stake (Sect.~\ref{sec:n-conclusion}).

\runin{Why a guarantee and a test complement each other.}
Method validation in Sect.~\ref{sec:n-discussion} says that a formal guarantee gives an upper limit on leakage and a validated test a lower one, and that for large models a strict limit often costs too much accuracy.
The cost comes from the noise that a small $\varepsilon$ requires, which in machine learning can be significant, and NIST encourages organisations to evaluate the risk of the final release as well, especially for values above 1~\cite[Sects.~2.2 and~3.5]{nist800226}.

A loose limit can still overstate the leakage.
The standard method for training with differential privacy adds noise at every training step~\cite{abadi2016dpsgd}.
Its guarantee is calculated for an attacker who sees the model after every step~\cite{nasr2021adversary}.
It then holds for the final model as well, since the final model is computed from those steps and a differential privacy guarantee is not weakened by further processing of what has been released, a property known as post-processing~\cite[Sect.~2.1]{nist800226}.
The final model can therefore leak at most as much as the limit says, and may leak less, by an amount that cannot be said in general.
Nasr et al.\ find that the limit is nearly reached by an attacker who has every capability the calculation allows, including choosing the whole dataset and altering the training steps, that seeing only the final model made little difference in their experiments, and that attacks are much weaker when the attacker cannot choose the data or interfere with training~\cite{nasr2021adversary}.
Annamalai constructs an artificial loss function for which the final model alone leaks as much as all the steps together, so a tighter limit for the final model cannot be proved for every training setup~\cite{annamalai2024loss}.
The limit can thus be conservative for a release, as it is a worst case over datasets, records and what the attacker can do, which is what Sect.~\ref{sec:n-discussion} means by ``under realistic assumptions''.

Audits, in which a model trained with a guarantee is attacked, have been developed to assess how tight the calculated limit is and to detect errors~\cite{steinke2023one}.
An audit states its result in the same terms as the guarantee, as a lower limit on $\varepsilon$, so the two can be compared directly~\cite{jagielski2020auditing,nasr2023tight,steinke2023one}.
The lower limit comes with a statistical confidence level~\cite{zanella2023bayesian}.
The distance between the two limits indicates how much room the guarantee leaves, as far as the audit uses a strong attack.

An audit also checks the implementation, on which the guarantee depends: if the audit finds more leakage than the guarantee permits, the implementation or the calculation is wrong.
NIST notes that auditing is useful for finding such errors but cannot prove that a guarantee holds~\cite[Sect.~2.7]{nist800226}.
An audit is itself a test, and needs validation as any other (Sect.~\ref{sec:n-discussion}).

\end{document}